\documentclass{article}

\usepackage[english]{babel}
\usepackage[utf8x]{inputenc}
\usepackage[T1]{fontenc}

\usepackage[a4paper,top=3cm,bottom=2cm,left=3cm,right=3cm,marginparwidth=1.75cm]{geometry}

\usepackage{amsmath}
\usepackage{graphicx}
\usepackage[colorinlistoftodos]{todonotes}
\usepackage[colorlinks=true, allcolors=blue]{hyperref}

\title{The Missing Mass Problem as a Manifestation of GR Contextuality}
\author{
W.M. Stuckey\footnote{Department of Physics, Elizabethtown College, Elizabethtown, PA 17022}, 
Timothy McDevitt\footnote{Department of Mathematics, Elizabethtown College, Elizabethtown, PA 17022}, 
A.K. Sten\footnote{Department of Physics, Elizabethtown College, Elizabethtown, PA 17022}, 
and Michael Silberstein\footnote{Department of Philosophy, Elizabethtown College, Elizabethtown, PA 17022 and Department of Philosophy, University of Maryland, College Park, MD  20742} 
}

\begin{document}
\maketitle

\begin{abstract}
In Newtonian gravity, mass is an intrinsic property of matter while in general relativity (GR), mass is a contextual property of matter, i.e., matter can simultaneously possess two different values of mass when it is responsible for two different spatiotemporal geometries. Herein, we explore the possibility that the astrophysical missing mass attributed to non-baryonic dark matter (DM) actually obtains because we have been assuming the Newtonian view of mass rather than the GR view. Since an exact GR solution for realistic astrophysical situations is not feasible, we explore GR-motivated ansatzes relating proper mass and dynamic mass for one and the same baryonic matter, as justified by GR contextuality. We consider four GR alternatives and find that the GR ansatz motivated by metric perturbation theory works well in fitting galactic rotation curves (THINGS data), the mass profiles of X-ray clusters (ROSAT and ASCA data) and the angular power spectrum of the cosmic microwave background (CMB, Planck 2015 data) without DM. We compare our galactic rotation curve fits to modified Newtonian dynamics (MOND), Burkett halo DM and Navarro-Frenk-White (NFW) halo DM. We compare our X-ray cluster mass profile fits to metric skew-tensor gravity (MSTG) and core-modified NFW DM. We compare our CMB angular power spectrum fit to scalar-tensor-vector gravity (STVG) and $\Lambda$CDM. Overall, we find our fits to be comparable to those of MOND, MSTG, STVG, $\Lambda$CDM, Burkett, and NFW. We present and discuss correlations and trends for the best fit values of our fitting parameters. For the most part, the correlations are consistent with well-established results at all scales, which is perhaps surprising given the simple functional form of the GR ansatz.
\end{abstract}

\section{Introduction}\label{sec1}
Since the early 1930's, galactic rotation curves (RC's) and galactic cluster masses have been known to deviate from Newtonian expectations based on luminous matter and known mass-to-luminosity ratios \cite{oort, zwicky1, zwicky2, rubin}. These two aspects of the missing mass problem \cite{garrett} make it one of the most persistent problems in physics. No one disputes the existence of baryonic dark matter, e.g., brown dwarfs, black holes, and molecular hydrogen, but there is wide agreement that baryonic dark matter does not exist in large enough supply to resolve the missing mass problem \cite{bergstrom}. As a consequence, the most popular attempt to resolve the missing mass problem is to assume the existence of some exotic new matter \cite{munoz, feng}. While the assumption of non-baryonic dark matter (DM) in $\Lambda$CDM cosmology works well for explaining cosmological features (scales greater than 1 Mpc), there is still no independent verification of non-baryonic dark matter \cite{young}, galactic RC's do not conform to the theoretical predictions of $\Lambda$CDM for the distribution of DM on galactic scales \cite{gentile1, deblok, clifton, mcgaughgalaxy, wyse}, the large number of DM predicted dwarf satellite galaxies has not been observed \cite{wyse, trippe}, and as Trippe writes, ``Indeed, none of the relevant empirical scaling relations of galactic kinematics -- notably the Tully–Fisher and Faber–Jackson relations, the surface density–acceleration and mass discrepancy–acceleration relations, and Renzo's rule -- follow from dark matter models a priori'' \cite{trippe}. Further, there are five seemingly incompatible properties that must be satisfied by DM, i.e.,  dark, cold, abundant, stable, and dissipationless \cite{carrollDM}. ``As Tom Shanks once said, there are only two things wrong with $\Lambda$CDM: $\Lambda$ and CDM'' \cite{bull}. Therefore, it is reasonable at this stage to consider alternatives that deny the need for DM in resolving the missing mass problem.

Foremost among alternatives to DM is the modification of existing theories of gravity. There are many such attempts, e.g., modified Newtonian dynamics (MOND) \cite{milgrom1, milgrom2, sanders1} and relativistic counterparts \cite{bekenstein, sanders2, zlosnik, zhao, blanchet}, Moffat's modified gravity (MOG), i.e., metric skew-tensor gravity (MSTG) and scalar-tensor-vector gravity (STVG) \cite{moffat1, moffat2, moffat3, moffat4, moffat5, moffat6, moffat7, moffat8}, Renormalization Group corrected General Relativity \cite{rodrigues}, and nonlocal general relativity \cite{mashhoon}. Arguably, given the well-known problems with DM stated above and the many favorable data fits of modified gravity programs, these modified gravity approaches are no less reasonable than DM approaches. Indeed, given that the missing mass problem is longstanding, it doesn't seem likely that missing mass phenomena will be accounted for within existing physics, so the most reasonable approaches are either a new theory of gravity, a new form of matter, or some combination of the two. Thus, these approaches rightfully account for the majority of effort.

On the other hand, the the missing mass problem \textit{is} longstanding despite extensive work on possible new physics and ``for realistic distributions of matter in galaxies, we have neither analytic, nor numerical solutions to general relativity from which orbits can be predicted'' \cite{brownstein}. Indeed, it is not reasonable to expect an exact GR solution applicable to most missing mass phenomena. So, perhaps it is reasonable to consider GR deviations from Newtonian gravity to account for missing mass phenomena, even in such weak gravitational fields. Indeed, Cooperstock et al. used GR instead of Newtonian gravity in fitting galactic RC's and found that the non-luminous matter in galaxies ``is considerably more modest in extent than the DM extent claimed on the basis of Newtonian gravitational dynamics'' \cite{magalhaes, carrick, cooperstock}. Cooperstock also showed \cite{cooperstock2} that there is no Newtonian limit for the flat rotation curves of GR's axially-symmetric van Stockum solution. While there are dissenting opinions on this result \cite{cross,fuchs,korzynski,menzies,martila}, we want to propose another mechanism by which GR deviates from Newtonian gravity in order to account for missing mass phenomena without DM.

\begin{figure}
\begin{center}
\includegraphics [height = 60mm]{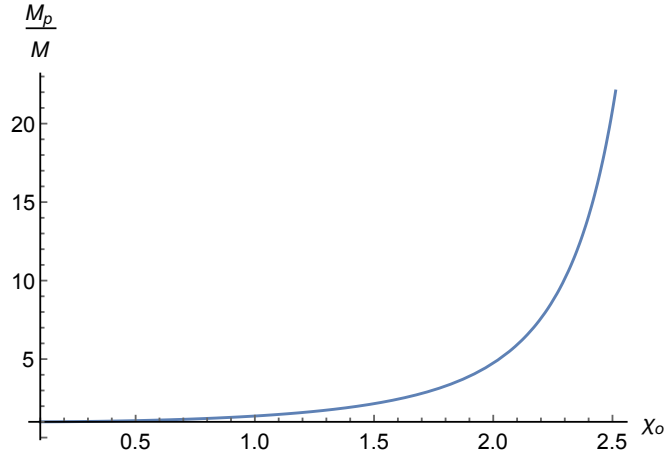}  \caption{Plot of $\frac{M_p}{M}$ as a function of the radial coordinate $\chi_o$ at the junction between the closed, matter-dominated FLRW spacetime and the Schwarzschild spacetime. In GR, unlike Newtonian gravity, matter can simultaneously have two different values of mass. Thus, mass is not an intrinsic property of matter in GR, it is contextual.} \label{Massratio}
\end{center}
\end{figure}

The difference we are pointing out is the well-known result per GR that matter can simultaneously possess two different values of mass when it is responsible for two different spatiotemporal geometries. This spatiotemporal contextuality of mass is not present in Newtonian gravity where mass is an intrinsic property of matter. For example, when a Schwarzschild vacuum surrounds a spherical matter distribution the ``proper mass'' $M_{p}$ of the matter, as measured locally in the matter, can be different than the ``dynamic mass\footnote{Typically, ``dynamical mass'' and ``luminous mass'' are the terms used with dynamical mass larger than luminous mass. Our terminology is following the GR convention.}'' $M$ in the Schwarzschild metric responsible for orbital kinematics about the matter \cite{wald}. This difference is attributed to binding energy and goes as $\displaystyle dM_p = \left(1-\frac{2GM(r)}{c^2r}\right)^{-1/2} \: dM$, which means it is too small to account for missing mass phenomena, but it is evidence of GR contextuality for mass. In another example, suppose a Schwarzschild vacuum surrounds a sphere of Friedmann-Lema{\^i}tre-Robertson-Walker (FLRW) dust connected at the instanteously null Schwarzschild radial coordinate. The dynamic mass $M$ of the surrounding Schwarzschild metric is related to the proper mass $M_{p}$ of the FLRW dust by \cite{stuckeyAJP} 
\begin{equation}
\frac {M_p}M = \left \{ 
{\renewcommand{\arraystretch}{1.75}%
\begin{tabular}{ll}
1 & flat model\\
$\displaystyle \frac{3 \left(\eta - \sin(\eta) \right)}{4\sin^{3}(\eta/2)} \geq 1$ & closed model \\
$\displaystyle \frac{3 \left(\sinh(\eta) - \eta \right)}{4\sinh^{3}(\eta/2)} \leq 1$ & open model
\end{tabular}
}
\right .
\label{massratio1}
\end{equation}
where $\eta$ is conformal time. Using this well-known embedding for the closed FLRW model (used originally to model stellar collapse \cite{misner}), we have for the ratio $\displaystyle \frac{M_p}{M}$ for a ball of FLRW closed-model dust surrounded by Schwarzschild vacuum as joined at FLRW radial coordinate $\chi_o$ 
\begin{equation}
\frac{M_p}{M} = \frac{3(2\chi_o -\sin(2\chi_o))}{4 \sin ^3(\chi_o)} \label{massratio}
\end{equation}
(Figure \ref{Massratio}), where
\begin{equation}
ds^2 = -c^2d\tau^2 + a^2(\tau)\left(d\chi^2 + \sin^2\chi d\Omega^2 \right) \label{FLRWmetric}
\end{equation}
is the closed FLRW metric. For $\displaystyle \chi_o = \frac{\pi}{2}$ Eq(\ref{massratio}) gives $\displaystyle \frac{M_p}{M} = 2.36$ and for $\displaystyle \chi_o = 0.8\pi$ Eq(\ref{massratio}) gives $\displaystyle \frac{M_p}{M} = 22.1$. The ratio quickly increases beyond $\displaystyle \chi_o = \frac{\pi}{2}$ because $dM < 0$, since the spherical area starts decreasing with increasing $\chi$. It is also the case that the extrinsic curvature of the interface changes sign for $\displaystyle \chi_o > \frac{\pi}{2}$, so this region has the dust surrounding the vacuum and is not what we're considering, but we're not proposing that this is an exact GR solution explaining away DM. Again, no such exact solution is likely forthcoming, given the complexity of matter distribution in most astrophysical situations. Rather, we're simply pointing out that GR contextuality allows for large differences between proper and dynamic mass. Essentially, mass is an intrinsic property of matter per Newtonian gravity, but mass is a geometric consequence of matter per GR. Since two different spacetime geometries may be associated with one and the same matter in a single GR solution, mass is a contextual property of matter per GR. We should quickly point out that this may prima facie seem to constitute a violation of the equivalence principle, as understood to mean inertial mass equals gravitational mass, since inertial mass can't be equal to two different values of gravitational mass. But, the equivalence principle says simply that spacetime is locally flat \cite{weinberg} and that is certainly not being violated here nor with any solution to Einstein's equations. So, we believe it is reasonable to consider this difference between GR and Newtonian gravity as the source of missing mass phenomena\footnote{We previously considered contextuality motivated by disordered locality requiring a modified gravity \cite{stuckeyIJMP2}, but here we consider the contextuality already inherent in GR.}. 

Consequently, in section \ref{sec2} we explore various options for a GR ansatz to obtain $M_p$ from $M$. Among the options, we find an ansatz that works well for galactic rotation curves, but fails for the mass profiles of X-ray clusters. We also find a GR ansatz that successfully fits galactic rotation curves (THINGS data) in section \ref{sec3}, the mass profiles of X-ray clusters (ROSAT and ASCA data) in section \ref{sec4}, and the angular power spectrum of the CMB (Planck 2015 data) in section \ref{sec5}, all without DM. We find our fits comparable to MOND, Burkett halo DM, NFW halo and core-modified DM, metric skew-tensor gravity (MSTG), scalar-tensor-vector gravity (STVG), and $\Lambda$CDM. We discuss the results and fitting factor trends in section \ref{sec6}. While we do not have enough data fits to draw strong conclusions per se, our results are consistent with well-established results elsewhere, so we feel justified in drawing speculative conclusions in section \ref{sec7}.

\section{The Ansatz}\label{sec2}
While proper mass $M_p$ must be $\sim 10$ times larger than dynamic mass $M$ in order to account for missing mass phenomena, the correction to the background spacetime metric responsible for missing mass phenomena is small. For example, assuming circular orbits (which is common for fitting galactic rotation curves) we have $\displaystyle v^2r = GM_p(r)$, where $M_p(r)$ is the (larger) proper mass inside the circular orbit at radius $r$ and $v$ is the orbital speed. This gives $\displaystyle \frac {2GM_p(r)}{c^2r}$ of the Schwarzschild metric equal to $\displaystyle 2\frac{v^2}{c^2}$. The largest galactic rotation speeds are typically only $10^{-3}c$, so the metric deviation from flat spacetime per the Schwarzschild metric is $\sim 10^{-6}$, an empirically small metric correction. 

Another way to quantify the difference geometrically between $M$ and $M_{p}$ is to use $\displaystyle ^{(3)}R = \frac {16 \pi G\rho}{c^2}$ \cite{wong} on galactic scales (where $\rho$ is the mass density in a thin spherical spatial annulus of a time and spherically symmetric spacetime solution). Assuming constant mass density (typical in a galactic bulge) out to radius $r$ where we have orbital velocity $v$, we have $\displaystyle \frac {16 \pi G\rho}{c^2} = \frac {12v^2}{r^2c^2}$. Constant density means $M(r) \sim r^3$ so
\begin{equation}
v(r) = \sqrt{\frac{GM(r)}{r}} \label{Eq7}
\end{equation}
increases linearly with distance from zero to its max value $v_{max}$. Assuming this happens at the radius of the bulge $r_b$ (galactic RC's actually peak farther out, so this overestimates the effect), we have $\displaystyle ^{(3)}R = \frac {12 v_{max}^2}{r_b^2c^2}$ for the spatial curvature scalar at $r$ rather than its Schwarzschild value of zero. $v_{max} \approx 300$ km/s and $r_b \approx 2000$ pc gives $^{(3)}R \sim 10^{-45}\: \mbox{m}^{-2}$. So, a factor of ten one way or the other in $\rho$ is geometrically inconsequential in this context. Intracluster medium (ICM) gas is even more rarefied and the potentials used for FLRW metric perturbations leading to anisotropies in the angular power spectrum of the CMB are already $\ll 1$ to include DM. 

In these arguments and throughout the paper, we have adopted the typical simplifying assumption of spherical symmetry in our missing mass phenomena. For galactic rotation curves it might seem more appropriate to use axisymmetry, so although ``The Kerr metric does not represent the exterior metric of a physically likely source, nor the metric during any realistic gravitational collapse'' \cite{teuk}, we should nonetheless consider it structurally. In Boyer-Linquist coordinates, the Kerr metric is (geometrized units)
\begin{equation}
ds^2 = -\left(1 - \frac {2Mr}{\Sigma} \right)dt^{2} - \frac {4Mar\sin^2\theta}{\Sigma}dtd\phi + \frac {\Sigma}{\Delta} dr^2 + \Sigma d\theta ^{2} + r^2 \left(1 + \frac {a^2}{r^2} + \frac {2Ma^2\mbox{sin}^2\theta}{r\Sigma} \right)d\phi ^2 \label{Eq4}
\end{equation}
where $\Sigma = r^2 + a^2cos^2\theta$ and $\Delta = r^2 - 2Mr + a^2$. This reduces to the Schwarzschild metric 
\begin{equation}
ds^2 = -\left(1 - \frac {2M}{r} \right)dt^{2} + \left(1 - \frac {2M}{r} \right)^{-1} dr^2 + r^2 d\Omega ^2 \label{Eq5}
\end{equation}
when $a \ll r$ ($\displaystyle a = \frac{J}{M}$ is the angular momentum per unit mass). If for a given galactic orbital radius $r$ we assume all the mass $M$ interior to that radius resides in a radially and axially thin, disk-shaped annulus at $r$ with orbital velocity $v$ in order to generate the most conservative (largest) estimate of $J$, we have $\displaystyle a = \frac {Mr^2 \omega} {Mc} = r \frac {v}{c}$ (restoring $c$). Since the largest galactic orbital velocities are about 300 km/s, we have $a \sim r/1000$ so we will use the much simpler, spherically symmetric Schwarzschild metric for our discussions of galactic rotation curves.

In summary, the missing mass phenomena we consider herein will be viewed geometrically and contextually per GR rather than dynamically and intrinsically per Newtonian gravity\footnote{This doesn't mean we will eschew dynamical discussion altogether. On the contrary, we will take a (3+1) approach for both our GR ansatz and the corresponding heuristic.}. In this geometric view, missing mass phenomena are understood as small deviations from some background metric. Of course, if one favors an exotic new kind of matter a la DM to account for missing mass phenomena, then it is important to note that the exotic new matter is far more prevalent than ordinary matter. But in our approach, it is more important to note that we're dealing with weak gravitational fields. It is customary to expect deviations of GR from Newtonian gravity for a strong gravitational field with its large spacetime curvature, but attempting to account for missing mass phenomena via GR actually requires that GR deviate from Newtonian gravity for weak fields, as we showed above. Thus, contrary to conventional thinking, what we're advocating is a geometric view of even weak gravitational fields, at least when the matter is distributed on astronomical scales. Of course, this is not unprecedented. 

In addition to Cooperstock et al. noted above, Moffat \& Rahvar used ``The weak field approximation of MOG'' as a perturbation of ``the metric and the fields around Minkowski space-time'' in fitting galactic RC's \cite{moffat6} and X-ray cluster mass profiles \cite{moffat7} without DM. Certainly MOND can be viewed in this fashion, since MOND advocates an extremely small modification to Newtonian acceleration on astronomical scales in the context of flat spacetime (Newtonian gravity), and acceleration due to gravity in flat spacetime is replaced by curved spacetime in GR. So, MOND could be viewed as saying that a small change in spacetime curvature (equivalent to a small change to acceleration in Newtonian gravity) replaces the need for a greatly increased mass in accounting for galactic dynamics. In fact, McGaugh et al. found \cite{mcgaughDM} ``a strong relation between the observed radial acceleration $g_{obs}$ and that due to the baryons, $g_{bar}$'' which ``suggests that the baryons are the source of the gravitational potential.'' And, as Trippe notes, ``the very existence of an empirical surface density–acceleration relation rules out the possibility that disk galaxy dynamics is dominated by dark matter; if this were the case, we would not expect a correlation between \textit{baryonic} surface density and acceleration at all'' \cite{trippe}. Thus, MOND could be interpreted as saying precisely what we're claiming, i.e., small spacetime curvature on astronomical scales can account for missing mass phenomena without DM. Simply put, that is our proposal.

Since 
\begin{itemize}
\item $M = M_p$ for the spatially flat FLRW model surrounded by Schwarzschild vacuum per Eq(\ref{massratio1}).
\item $^{(3)}R = 0$ for both the spatially flat FLRW dust and Schwarzschild vacuum.
\item We're talking about weak gravitational fields.
\end{itemize}
we might attribute the large mass difference between $M$ and $M_p$ to a small difference in \textit{spatial} curvature. Indeed, the difference between $M_p$ and $M$ shown in Eq(\ref{massratio}) obtains from the integrated difference in spatial geometry between the closed, matter-dominated FLRW dust ball and the surrounding Schwarzschild vacuum \cite{stuckeyAJP}
\begin{equation}
\frac{M_p}{M} = \frac{\int_{0}^{\chi_o}\sin^2\chi \: d\chi}{\int_{0}^{\chi_o}\sin^2\chi \: d(\sin\chi)} \label{integratedmassratio}
\end{equation}
Since in all three cases (galaxies, clusters, and CMB perturbations) we are assuming spherical symmetry, this (3+1) view is not inconsistent with Birkhoff's theorem, i.e., any spherically symmetric vacuum solution of GR will be static. Accordingly, consider the following heuristic. 

Suppose that the Schwarzschild vacuum surrounding the FLRW dust ball in our example above is itself surrounded by the remaining FLRW dust, i.e., the ball of FLRW dust has collapsed out of its FLRW cosmological context and is now separated from that cosmological context by the Schwarzschild vacuum. The spacetime geometry of the surrounding FLRW dust will be unaffected by the intervening Schwarzschild vacuum, so observers in the surrounding FLRW dust (global context) will obtain the ``globally determined'' proper mass $M_p$ for the collapsed dust ball while observers in the Schwarzchild vacuum (local context) will obtain the ``locally determined'' dynamic mass $M$ for the collapsed dust ball. Thus, stellar mass-to-luminosity ratios would be based on $M$ and would not give the proper mass $M_p$ required to explain galactic RC's. Essentially, per this heuristic, we imagine that coalescing matter sweeps up very small amounts of spatial curvature, leaving flatter empty space in its wake. While it might seem from this heuristic that we're merely replacing DM with spatial curvature a la $\displaystyle ^{(3)}R = \frac {16 \pi G\rho}{c^2}$ \cite{wong} there is a major difference, i.e., in the Newtonian view, observers inside the ball of dust and observers in the surrounding vacuum would all measure the same increased value of mass due to the addition of spatial curvature. In the GR view, those observers measure two different values for the mass of the ball. The real underlying explanation for missing mass phenomena here is GR contextuality, not the mere addition of swept-up spatial curvature. Even though the amount of spatial curvature swept up is small, the resulting difference between $M$ and $M_p$ is large. Of course, this is an idealization and the actual situation in a galaxy would be far more complicated since such a nested solution would have to marry up with other such nested solutions. Indeed, the surrounding FLRW dust may not even remain, having coalesced into other bodies and clouds. And, as we will see, the largest contribution to the correction of galactic dynamic mass is not from the stellar disk or bulge, but the gas. Again, given the complexity of GR, no such exact solution can be expected, so some approximation method for obtaining the proper mass from the dynamic mass must be motivated and checked for efficacy against astrophysical data. Here we will consider four options, beginning with an acceleration option.

In presenting MOND's ansatz for a correction to Newtonian acceleration, Trippe notes, ``Multiple observations show that the discrepancy between observed and dynamical mass is a function of gravitational acceleration (or field strength) but not of other parameters (size, rotation speed, etc.) of a galaxy'' \cite{trippe}. So, it is perhaps advisable to compare the GR accelerations for constant ``Schwarzschild $r$'' observers in the closed, matter-dominated FLRW model and the Schwarzschild vacuum. In the Schwarzschild vacuum, the acceleration (rate of deviation from geodetic motion) is purely radial and this component is given by $\displaystyle \frac{GM}{r^2}$ outward. This looks deceptively Newtonian, given the difference in the concept of acceleration, i.e., in Newtonian physics the acceleration is zero for this observer and is $\displaystyle \frac{GM}{r^2}$ inwards for the observer in free fall while in GR the acceleration of the free fall observer is zero, so the constant $r$ observer is accelerating radially outward. The difference is that the magnitude of this GR acceleration is 
\begin{equation}
\mid a_r \mid = \frac{1}{\sqrt{1-\frac{2GM}{c^2r}}}\frac{GM}{r^2}
\end{equation}
where $M$ is the dynamic mass. The acceleration for the constant ``Schwarzschild $r$'' $= a(\eta)\sin{\chi}$ observer in the closed, matter-dominated FLRW expanding dust has both radial ($\chi$) and temporal components. Here $\displaystyle a(\eta) = \frac{B}{2}\left(1 - \cos{\eta}\right)$ is the scaling parameter for the comoving coordinate system in Eq(\ref{FLRWmetric}) with $a(\eta)d\eta = d\tau$ where $\tau$ is the proper time for comoving FLRW observers (units of $c = 1$). The magnitude of the radial component of the FLRW constant ``Schwarzschild $r$'' observer's acceleration $\mid a_{\chi}\mid \: = \sqrt{g_{\chi\chi}}\: a_{\chi}$ is much more complicated than its Schwarzschild counterpart
\begin{equation}
\mid a_{\chi}\mid \: = \frac{4 \sqrt{2} r \csc ^2\left(\frac{\eta }{2}\right) \sqrt{-4 B^2 \cos (\eta )+B^2 \cos (2 \eta )+3 B^2-8 r^2} \: \beta}{B \left(15 B^2 \cos (\eta )-6 B^2 \cos (2 \eta )+B^2 \cos (3 \eta )-10 B^2+32 r^2\right)^2} \label{GRacceleration}
\end{equation}
where
\begin{equation}
\beta = 6 B^2 \cos (2 \eta )-B^2 \cos (3 \eta )+\left(48 r^2-15 B^2\right) \cos (\eta )+10 B^2+16 r^2
\end{equation}
and
\begin{equation}
B = \frac{4a^2 M_p}{3\left(a^2 \arcsin{\left(\frac{r}{a}\right)} - r \sqrt{a^2 - r^2}\right)}
\end{equation}
Therefore, even in this simple exact solution the geometric difference between dynamic mass $M$ and proper mass $M_p$ in the context of acceleration is not transparent, so we seek a simpler approach to our GR ansatz.

We might look for a functional form using Eq(\ref{integratedmassratio}) according to which $dM_p = \frac{dM}{\cos \chi}$. In fact, using 
\begin{equation}
dM_p = \left(B + \frac{A}{\cos \left(\frac{\pi r}{4 r_{max}}\right)}\right)dM \label{GRansatz1}
\end{equation}
produced a slightly better fit of THINGS data than the best GR ansatz for both X-ray clusters and galactic RC's. As stated in Table \ref{GRRC1}, this fit produced an average mean square error (MSE) of 81.9 $(\mbox{km/s})^2$ versus 101 $(\mbox{km/s})^2$ from Table \ref{GRRC} results for the overall best GR ansatz fit. However, this form failed to produce comparable fits for the X-ray cluster mass profiles. For example, the best GR ansatz fit (see below) produced an MSE of 0.00173 (from $(\Delta \mbox{Log}(M))^2$, see Table \ref{GRclusters} and Figure \ref{cluster2}) for the Perseus cluster while Eq(\ref{GRansatz1}) produced an MSE of 0.114 (see Figure \ref{Perseus}) for this cluster. Variations on this fitting function faired no better, so we might consider a fitting function based on the functional form for binding energy, i.e., $dM_p = \sqrt{g_{rr}}\:dM$. This form would be something like
\begin{equation}
dM_p = \frac{dM}{\sqrt{B - \frac{A}{r}}} \label{GRansatz2}
\end{equation}
The problem with this form is that it necessarily becomes imaginary for finite $B$, since we must in all cases consider $r \rightarrow 0$. For binding energy this is avoided since $A(r)$ is such that $\frac{A}{r} \rightarrow 0$ as $r \rightarrow 0$, but we desire a simple $A$ for fitting purposes. Thus, in searching for a GR-motivated ansatz, we've eliminated GR acceleration differences for constant ``Schwarzschild $r$'' observers in Schwarzschild and FLRW closed spacetimes, proper versus dynamic mass differences for those same observers, and proper versus dynamic mass differences as motivated by GR binding energy. Certainly, one may find other forms for a GR-motivated ansatz, but let us reveal one we found to work well for all three missing mass phenomena herein.

In the case of dark energy as pertains to the SCP Union2.1 SN Ia data, we considered metric corrections $h_{\alpha \beta}$ to proper distance $D_p$ satisfying the vacuum perturbation equation $\displaystyle \nabla^2 h_{\alpha \beta} = 0$ in the flat space of FLRW matter-dominated cosmology, i.e., $\displaystyle \frac{d^2}{d D_p ^2} h_{ii} = 0$ where $D_p$ is proper distance per the FLRW metric \cite{stuckeyCQG,stuckeyIJMP1}. We then corrected proper distance according to $\displaystyle D_p \rightarrow \sqrt{1 + h_{ii}} \: D_p = \sqrt{1 + \frac{D_p}{A}} \: D_p$ ($A$ is an arbitrary constant used as a fitting parameter and was found to be about 8 Gcy). Here we adopt this approach for missing mass phenomena. We will assume $\displaystyle dM_p = \sqrt{1 + h} \: dM$ in analogy with our correction of proper distance above. Further, $\nabla^2 h = 0$ with spherical symmetry assumed for galactic rotation curves, X-ray cluster mass profiles, and the baryon-photon perturbations in pre-recombination FLRW cosmology. The important difference between our treatment of dark energy and missing mass is that while $h$ satisfies the perturbation equation, it does not have to be the case that $h \ll 1$. But, as explained above, it is the case that even the enhanced proper mass constitutes a small perturbative correction to a background metric, so the motivation for this GR ansatz isn't totally ill-founded. Accordingly, the GR ansatz is based on $\displaystyle h = \frac{A}{r} + B$ which is used to obtain proper mass from dynamic mass per
\begin{equation}
dM_p = \sqrt{1 + h} \: dM = \sqrt{\frac{A}{r} + B} \: dM \label{GRansatz}
\end{equation}
As with dark energy, the arbitrary constants $A$ and $B$ are used as fitting parameters. The best fit values of $A$ and $B$ show interesting trends across and within the three data sets that we will discuss in section \ref{sec6}. We now proceed to the fits.

\section{Galactic Rotation Curves}\label{sec3}
In the MOND approach with spherical symmetry and circular orbits, as is commonly used for fitting galactic RC's, it is the Newtonian acceleration
\begin{equation}
a_N(r) = \frac{GM(r)}{r^2} \label{Eq8}
\end{equation}
that is modified by the so-called interpolating function $\mu(x)$ where $\displaystyle x = \left| \frac {a(r)}{a_o} \right|$ with $a_o$ a universal constant so that
\begin{equation}
\mu(x)a(r) = \frac{GM(r)}{r^2} \label{Eq9}
\end{equation}
$\mu(x) \sim x$ for $x \ll 1$ and $\mu(x) \sim 1$ for $x \gg 1$.
Of course, $\mu(x(r))$ means one could view Eq(\ref{Eq9}) as a modification to $M(r)$ or $G$ as in MSTG \cite{brownstein} 
\begin{equation}
a(r) = \frac{GM(r)}{\mu(r)r^2} \label{Eq10}
\end{equation}
where $\displaystyle G(r) = \frac {G}{\mu(r)}$ in MSTG. In STVG in $k$ space we also have $G \rightarrow G_{eff}$, as we will see in section \ref{sec5}. In terms of the Newtonian potential $\chi$, the Schwarzschild metric is 
\begin{equation}
ds^2 = -c^{2} \left(1 + \frac {2 \chi}{c^2} \right)dt^{2} + \left(1 + \frac {2 \chi}{c^2} \right)^{-1} dr^2 + r^2 d\Omega ^2 \label{Eq11}
\end{equation}
where ${a(r)} = \left|\nabla(\chi)\right|$. So, in that sense MOND, MSTG, and STVG can all be viewed as MOG \cite{brownstein}. However, our goal here is not to critique these well-established programs, but to use them as standards. For example, we will test our GR ansatzes in fitting galactic RC's by comparing them to MOND using data where MOND fits were deemed ``very successful'' \cite{gentile}. We will also provide Burkett and NFW DM halo fits of the THINGS data to bracket our GR ansatzes with the two mainstream approaches, i.e., DM and modified gravity. We begin with the MOND fits.

In the MOND fits, we will use the so-called ``simple'' $\mu$ function
\begin{equation}
\mu(x) = \frac{x}{1+x} \label{Eq12} 
\end{equation}
which Gentile et al. found most successful with the THINGS data \cite{gentile}. We will also adopt their median best-fit value of the acceleration parameter for the simple $\mu$ function, $a_o = (1.22 \pm 0.33)$ x $10^{-8}$ cm s$^{-2}$. Generally speaking, MOND has a single fit parameter, $M/L$ for the stellar disk, although when bulge data are available independent of the stellar disk, an additional $M/L$ factor can be introduced for the bulge as well. We will also allow galactic distance $d$ to vary up to one sigma from the nominal value in our MOND fits. Gentile et al. called this ``$d$ constrained'' \cite{gentile}. Finally, Gentile et al. also note these data are ``not obviously dominated by non-circular motions,'' so we assume circular orbits to obtain the MOND $v$ as related to the Newtonian $v_N$ \cite{richtler}
\begin{equation}
v^2 = \frac{v_N^2}{2} + \sqrt{\frac{v_N^4}{4}+v_N^2 a_o r} \label{Eq13}
\end{equation}
where 
\begin{equation}
v_N^2 = v_{gas}^2 + v_{disk}^2 + v_{bulge}^2 \label{Eq14}
\end{equation}
Allowing for variation in $M/L$ for the stellar disk and bulge means 
\begin{equation}
v_N^2 = v_{gas}^2 + v_{disk}^2 \gamma_{disk} + v_{bulge}^2 \gamma_{bulge} \label{Eq15}
\end{equation}
where $\gamma_{disk}$ and $\gamma_{bulge}$ are fit parameters. Assuming the distance $d$ to the galaxy can vary means a further correction in $M/L$ (fitting parameter) for all components, call it $\gamma_{dist}$, giving
\begin{equation}
v_N^2 = \left(v_{gas}^2 + v_{disk}^2 \gamma_{disk} + v_{bulge}^2 \gamma_{bulge} \right) \gamma_{dist} \label{Eq16}
\end{equation}
where $r \rightarrow \gamma_{dist}r$ in Eq(\ref{Eq13}) as well. Eq(\ref{Eq13}) is then used to fit THINGS data, i.e., $v(r)$, $v_{gas}(r)$, $v_{bulge}(r)$, and $v_{disk}(r)$, using the fit parameters $\gamma_{disk}$, $\gamma_{gas}$, and $\gamma_{dist}$. 

The Burkett and NFW DM halo distributions are from Karukes \& Salucci \cite{karukes}. The galactic RC is decomposed as in MOND according to 
\begin{equation}
v^2(r) = v_{disk}^2(r) + v_{bulge}^2(r) + v_{gas}^2(r) + v_{DM}^2(r) \label{DMRC}
\end{equation}
In their DM fits, they account for the variation of $M/L$ in the disk data $v_{diskdata}$ by adding an additional fitting parameter (stellar mass-to-light ratio) $\Upsilon_{\ast}^{3.6}$ so that $\displaystyle v_{disk}^2 = \frac{\Upsilon_{\ast}^{3.6}}{0.57} v_{diskdata}^2$. The Burkett empirical density distribution of DM is given by
\begin{equation}
\rho_B(r) = \frac{\rho_o r_{c}^3}{\left(r + r_c \right)\left(r^2 + r_{c}^2 \right)}
\end{equation}
where $\rho_o$ is the central density and $r_c$ is the core radius. These and $\Upsilon_{\ast}^{3.6}$ are the three fitting parameters for Eq(\ref{DMRC}) where
\begin{equation}
v_{DM}^2(r) = \frac{4\pi \rho_o r_{c}^3 G}{r}\left( -\frac{1}{2}\arctan\left(\frac{r}{r_c}\right) + \frac{1}{2}\ln\left(\frac{r+r_c}{r_c}\right) + \frac{1}{4}\ln\left(\frac{r^2+r_{c}^2}{r_{c}^2}\right)\right)
\end{equation}
The NFW DM profile is given by
\begin{equation}
\rho_{NFW}(r) = \frac{\rho_s}{\left(\frac{r}{r_s}\right)\left(1 + \frac{r}{r_s}\right)^2}
\end{equation}
where $\rho_s$ is the characteristic density and $r_s$ is the scale radius of the DM distribution. Again, these and $\Upsilon_{\ast}^{3.6}$ are the three fitting parameters for Eq(\ref{DMRC}) where in this case
\begin{equation}
v_{DM}^2(r) = \frac{4\pi \rho_s r_{s}^3 G}{r}\left(\ln\left(\frac{r+r_s}{r_s}\right) - \frac{r}{r + r_s}\right)
\end{equation}

Our GR ansatz computes the baryonic dynamic mass $M_{i}(r)$ for each component $i$ using Eq(\ref{Eq7}) with $v_i(r)$ supplied by THINGS. Then $\Delta M_{i} = M_{i}(r_2) - M_{i}(r_1)$ for the dynamic mass of the $i^{th}$ component in an annulus. The geometric modification of the GR ansatz is applied to each $\Delta M_{i}$ according to its radial location to obtain the corresponding proper mass $\Delta M_{p_i}$
\begin{equation}
\Delta M_{p_i} = \sqrt{\frac{2A_i}{r_1 + r_2} + B} \; \Delta M_{i} \label{Eq17}
\end{equation}
where $A_i$ is a fitting factor for the $i^{th}$ component and $B$ is a fitting parameter that is the same for all components. Thus, since the dynamic mass is a locally determined quantity, the global proper mass requires an integration of the corrected dynamic mass, just as in computing binding energy \cite{wald}. The $\Delta M_{p_i}$ are summed to produce $M_{p_i}(r)$, which are then summed to produce $M_p(r)$ which gives $v(r)$ per Eq(\ref{Eq7}).

We are now ready to compare (via tables and figures) our GR ansatz Eqs(\ref{GRansatz1} \& \ref{GRansatz}) fits with MOND, Burkett, and NFW fits of twelve high-resolution galactic RC's from The HI Nearby Galaxy Survey \cite{things} used by Gentile et al. \cite{gentile} to explore MOND fits. [See Moffat \& Rahvar \cite{moffat6} for MOG fits of THINGS data. See Siutsou et al. for Burkett fits of THINGS data \cite{siutsou}.] Gentile et al. describe these data as ``the most reliable for mass modelling, and they are the highest quality RC's currently available for a sample of galaxies spanning a wide range of luminosities.'' Our GR ansatz results are in Tables \ref{GRRC1} and \ref{GRRC}, the MOND results are in Table \ref{mondRC}, the Burkett results are in Table \ref{BurkettRC}, and the NFW results are in Table \ref{NFWRC}. Graphical display of the best fits reported in these tables are in Figures \ref{galaxy1} \& \ref{galaxy2}. These results are discussed in section \ref{sec6} after obtaining results from X-ray cluster mass profiles (section \ref{sec4}) and anisotropies in the CMB angular power spectrum (section \ref{sec5}).

\begin{table}
\begin{center}
\begin{tabular}{c|ccccc}
Name & $A_{bulge}$ & $A_{disk}$ & $A_{gas}$ & $B$ & MSE \\
\hline
NGC 2841& 0.924 & 2.06 & 22.5 & 0.00 & 89.8 \\
\hline
NGC 7331& 0.011 & 0.825 & 4.24 & 0.169 & 53.0 \\
\hline
NGC 3521& N/A & 0.874 & 5.23 & 0.00 & 236 \\
\hline
NGC 6946& 0.0056 & 0.369 & 7.56 & 0.580 & 53.4 \\
\hline
NGC 2903& 0.00 & 3.83 & 13.4 & 0.00 & 259 \\
\hline
NGC 5055& 0.145 & 0.582 & 9.41 & 0.118 & 105 \\
\hline
NGC 3198& 0.00 & 1.19 & 4.81 & 0.361 & 35.3 \\
\hline
NGC 3621& N/A & 0.00 & 4.44 & 1.27 & 61.8 \\
\hline
NGC 2403& N/A & 2.35 & 9.01 & 0.00 & 33.2 \\
\hline
NGC 7793& N/A & 0.00 & 5.68 & 1.87 & 31.8 \\
\hline
NGC 2976& N/A & 0.908 & 3.51 & 0.00 & 22.3 \\
\hline
DDO 154& N/A & 9.18 & 5.01 & 0.00 & 1.87 \\
\hline
\end{tabular}
\end{center}
\caption{GR ansatz fits for THINGS using Eq(\ref{GRansatz1}). $A_i$ and $B$ are dimensionless and mean square error (MSE) is in $(\mbox{km/s})^2$. The fitting form used here worked slightly better than the GR ansatz shown in Table \ref{GRRC} (average MSE = 81.9 $(\mbox{km/s})^2$ versus 101 $(\mbox{km/s})^2$ from Table \ref{GRRC}), but it failed for X-ray cluster mass profiles (Figure \ref{Perseus}).}
\label{GRRC1}
\end{table}

\begin{table}
\begin{center}
\begin{tabular}{c|ccccc}
Name & $A_{bulge}$ & $A_{disk}$ & $A_{gas}$ & $B$ & MSE \\
\hline
NGC 2841& 0.00 & 19.8 & 17600 & 0.00 & 106 \\
\hline
NGC 7331& 0.00 & 0.00 & 645 & 0.67 & 63.5 \\
\hline
NGC 3521& N/A & 0.00 & 442 & 0.67 & 243 \\
\hline
NGC 6946& 0.00 & 0.00 & 1030 & 0.56 & 75.8 \\
\hline
NGC 2903& 0.00 & 18.5 & 3730 & 0.00 & 489 \\
\hline
NGC 5055& 0.00 & 1.16 & 2430 & 0.03 & 105 \\
\hline
NGC 3198& 0.00 & 8.33 & 705 & 0.014 & 10.1 \\
\hline
NGC 3621& N/A & 0.00 & 515 & 0.97 & 37.4 \\
\hline
NGC 2403& N/A & 4.69 & 922 & 0.00 & 29.9 \\
\hline
NGC 7793& N/A & 0.00 & 286 & 2.82 & 18.8 \\
\hline
NGC 2976& N/A & 0.00 & 39.3 & 0.794 & 23.9 \\
\hline
DDO 154& N/A & 0.00 & 100 & 0.00 & 6.44 \\
\hline
\end{tabular}
\end{center}
\caption{GR ansatz fits for THINGS using Eq(\ref{GRansatz}). $A_i$ are in kpc, $B$ is dimensionless, and mean square error (MSE) is in $(\mbox{km/s})^2$. Corresponding plots are in Figures \ref{galaxy1} \& \ref{galaxy2}.}
\label{GRRC}
\end{table}

\begin{table}
\begin{center}
\begin{tabular}{c|cccc}
Name & $\gamma_{bulge}$* & $\gamma_{disk}$ & $\gamma_{dist}$ & MSE \\
\hline
NGC 2841& 0.99 (1.02) & 0.79 (0.70) & 1.46 (1.53) & 19.5 \\
$d$ constrained& 1.09 (1.24) & 1.32 (1.20) & 1.11 (1.11) & 54.1 \\
\hline
NGC 7331& 1.09 (1.16) & 0.98 (0.91) & 0.68 (0.71) & 18.8  \\
$d$ constrained& 1.17 (1.22) & 0.59 (0.57) & 0.913 (0.913) & 28.6  \\
\hline
NGC 3521& N/A & 1.77 (1.08) & 0.44 (0.65) & 210 \\
$d$ constrained& N/A & 0.93 (0.97) & 0.70 (0.70) & 264 \\
\hline
NGC 6946& 0.58 (0.55) & 0.94 (0.58) & 0.80 (1.09) & 29.2 \\
comparison**& 0.49 (0.55) & 0.59 (0.58) & 1.09 (1.09) & 32.3 \\
\hline
NGC 2903& 0.00 (0.00) & 2.06 (2.80) & 1.25 (1.07) & 202 \\
\hline
NGC 5055& 0.40 (0.47) & 0.80 (0.84) & 0.66 (0.62) & 26.7 \\
$d$ constrained& 0.38 (0.43) & 0.72 (0.70) & 0.70 (0.70) & 27.2 \\
\hline
NGC 3198***& 0.37 (N/A) & 1.59 (N/A) & 0.58 (0.62) & 18.8 \\
$d$ constrained& 0.26 (N/A) & 0.65 (N/A) & 0.89 (0.89) & 74.2 \\
\hline
NGC 3621& N/A & 0.76 (0.75) & 0.91 (0.92) & 9.14 \\
\hline
NGC 2403& N/A & 0.69 (0.63) & 1.53 (1.46) & 34.4 \\
$d$ constrained& N/A & 1.29 (1.10) & 1.08 (1.08) & 42.3 \\
\hline
NGC 7793& N/A & 0.51 (0.39) & 1.52 (1.68) & 42.1 \\
$d$ constrained& N/A & 1.00 (0.90) & 1.10 (1.10) & 55.0 \\
\hline
NGC 2976& N/A & 0.11 (0.09) & 2.02 (2.17) & 7.83 \\
$d$ constrained& N/A & 0.38 (0.36) & 1.10 (1.10) & 12.0 \\
\hline
DDO 154& N/A & 4.01 (2.47) & 0.64 (0.71) & 1.63 \\
$d$ constrained& N/A & 1.36 (1.56) & 0.75 (0.75) & 4.60 \\
\hline
\end{tabular}
\end{center}
\caption{MOND fits for THINGS. $\gamma_i$ are dimensionless and mean square error (MSE) is in $(\mbox{km/s})^2$. Values in parentheses are from Gentile et al. fits minimizing reduced $\chi^2$ \cite{gentile}. If there is no ``$d$ constrained,'' the ``$d$ free'' fit satisfies ``$d$ constrained.'' Corresponding plots are in Figures \ref{galaxy1} \& \ref{galaxy2}. *N/A means Gentile et al. did not use independent bulge data. **For comparison with Gentile et al. we fixed $\gamma_{dist} = 1.09$. ***Gentile et al. used composite disk and bulge data in THINGS, while we used the separate disk and bulge data also supplied in THINGS.} \label{mondRC}
\end{table}

\begin{table}
\begin{center}
\begin{tabular}{c|cccc}
Name & $\rho_{o}$ & $r_{c}$ & $\Upsilon_{\ast}^{3.6}$ & MSE \\
\hline
NGC 2841& 9.71 (3.52) & 18.3 (20.6) & 1.00 (1.44) & 41.8 (1035) \\
\hline
NGC 7331& 11.3 (16.9) & 12.3 (9.70) & 0.395 (0.378) & 24.8 (26.8) \\
\hline
NGC 3521& 845 (40.6) & 1.92 (5.40) & 0.00 (0.395) & 91.2 (168) \\
\hline
NGC 6946& 153 (98.9) & 3.75 (3.48) & 0.038 (0.356) & 51.6 (179) \\
\hline
NGC 2903& 138 (263) & 3.80 (2.89) & 0.606 (0.264) & 522 (548) \\
\hline
NGC 5055& 58.1 (5.42) & 5.81 (14.3) & 0.00 (0.289) & 541 (736) \\
\hline
NGC 3198& 7.25 (24.4) & 10.9 (6.32) & 0.571 (.319) & 72.9 (80.1) \\
\hline
NGC 3621& 7.47 (23.2) & 11.5 (6.48) & 0.616 (0.413) & 13.2 (38.0) \\
\hline
NGC 2403& 12.1 (56.2) & 8.89 (4.08) & 1.28 (0.543) & 28.5 (74.1) \\
\hline
NGC 7793& 66.3 (88.0) & 2.92 (2.60) & 0.568 (0.424) & 31.0 (32.6) \\
\hline
NGC 2976& 69.4 (27.1) & 2.40 (20.0) & 0.284 (0.375) & 5.54 (7.09) \\
\hline
DDO 154& 9.56 (19.6) & 3.37 (2.32) & 5.25 (1.06) & 1.14 (1.78) \\
\hline
\end{tabular}
\end{center}
\caption{Burkett DM halo fits for THINGS. $\rho_o$ is in $10^{-25}$ g cm$^{-3}$, $r_c$ is in kpc, $\Upsilon_{\ast}^{3.6}$ is dimensionless, and mean square error (MSE) is in $(\mbox{km/s})^2$. $\rho_o$ and $r_c$ values in parentheses are from Siutsou et al. Burkett DM fits of THINGS minimizing reduced $\chi^2$ \cite{siutsou}. $\Upsilon_{\ast}^{3.6}$ values in parentheses are then our best fit values using Siutsou $\rho_o$ and $r_c$ values resulting in MSE values in parentheses.} \label{BurkettRC}
\end{table}

\begin{table}
\begin{center}
\begin{tabular}{c|cccc}
Name & $\rho_{s}$ & $r_{s}$ & $\Upsilon_{\ast}^{3.6}$ & MSE \\
\hline
NGC 2841& 6.29 & 23.3 & 0.727 & 29.8 \\
\hline
NGC 7331& 1.99 & 33.5 & 0.349 & 23.2  \\
\hline
NGC 3521& 15.8 & 8.99 & 0.354 & 191 \\
\hline
NGC 6946& 1.23 & 46.3 & 0.394 & 29.2 \\
\hline
NGC 2903& 5.46 & 17.7 & 1.36 & 662 \\
\hline
NGC 5055& 7.67 & 14.8 & 0.0673 & 666 \\
\hline
NGC 3198& 0.99 (0.99) & 31.5 (32.9) & 0.546 (0.80) & 101 \\
\hline
NGC 3621& 16.2 & 3.30 & 0.444 & 3.25 \\
\hline
NGC 2403& 21.1 & 6.75 & 0.115 & 16.2 \\
\hline
NGC 7793& 14.2 & 7.22 & 0.175 & 49.6 \\
\hline
NGC 2976& 0.0819 & 573 & 0.200 & 9.96 \\
\hline
DDO 154& 1.54 & 9.57 & 0.167 & 2.89 \\
\hline
\end{tabular}
\end{center}
\caption{NFW DM halo fits for THINGS. $\rho_s$ is in $10^{-25}$ g cm$^{-3}$, $r_s$ is in kpc, $\Upsilon_{\ast}^{3.6}$ is dimensionless, and mean square error (MSE) is in $(\mbox{km/s})^2$. Values in parentheses are from Karukes \& Salucci \cite{karukes} fit of HALOGAS survey data \cite{halogas} for NGC 3198. They described their $\rho_s$ value as ``too low'' and their $r_s$ value as ``too high.''} \label{NFWRC}
\end{table}
 
\section{X-ray Cluster Mass Profiles}\label{sec4}
For fitting the mass profiles of X-ray clusters, one assumes a spherically symmetric, isothermal distribution of X-ray emitting gas in hydrostatic equilibrium with the gravitational potential \cite{moffat2}. The gas mass profile as determined by the X-ray luminosity is given by 
\begin{equation}
M(r) = 4 \pi \int_{0}^{r} \rho(r')r'^2 dr'\label{Eq18}
\end{equation}
where 
\begin{equation}
\rho(r) = \rho_o \left(1 + \left(\frac{r}{r_c}\right)^2 \right)^{-1.5\beta}. \label{Eq19}
\end{equation}
according to the King $\beta$-model. The constants $\rho_o$, $r_c$ and $\beta$ are found from fits of the surface density map $\Sigma(x,y)$. $M(r)$ corresponds to the ``dynamic mass'' for galactic RC's above. The proper mass profile needed for hydrostatic equilibrium at temperature $T$ is given by 
\begin{equation}
M_p(r) = \frac{3\beta k T}{G \mu_A m_p}\left(\frac{r^3}{r^2 + r_c^2}\right) \label{Eq20}
\end{equation}
where $\mu_A$ is the mean atomic weight of the gas constituents ($\approx 0.609$), $m_p$ is the mass of the proton, $k$ is the Boltzmann constant, and $G$ is the gravitational constant. In the MSTG version of MOG per Moffat \& Brownstein, as stated above, $G \rightarrow G(r)$ and is understood to reduce $M_p(r)$ of Eq(\ref{Eq20}) to $M(r)$ of Eq(\ref{Eq18}). Effectively, we have
\begin{equation}
M(r) = M_p(r) + M_o\eta(r) - \sqrt{(M_o\eta(r))^2 + 2M_oM(r)\eta(r)} \label{Eq21} 
\end{equation}
where
\begin{equation}
\eta(r) = 0.5\left(1 - \exp(-r/r_o)\left(1 + \frac{r}{r_o}\right)\right)^2 \label{Eq22}
\end{equation}
with $M_o$ and $r_o$ fitting parameters and $M_p(r)$ computed using $G$ in Eq(\ref{Eq20}). 

According to the DM hypothesis, there is an additional DM contribution to $\rho(r')$ in Eq(\ref{Eq18}) that leads to $M_p$ of Eq(\ref{Eq20}). Here we use core-modified NFW as in Brownstein \cite{brownstein} for that contribution
\begin{equation}
\rho_{DM}(r) = \frac{\rho_o r_{s}^3}{\left(r^3 + r_{s}^3 \right)}
\end{equation}
where $\rho_o$ is the central density and $r_s$ is the scaling radius. 

Since we're assuming spherical symmetry, our approach used for galactic RC's can be easily adopted here. As with the DM fit and contrary to MSTG, we are increasing $dM(r)$ by a geometric factor to give the proper mass $M_p(r)$. Thus, we modify Eq(\ref{Eq18}) to give
\begin{equation}
M_p(r) = \int_{0}^{r} \sqrt{B + \frac{A}{r'}} \: dM = 4 \pi\int_{0}^{r} \sqrt{B + \frac{A}{r'}} \: \rho(r')r'^2 dr' \label{Eq23}
\end{equation}
with $\rho(r)$ given by Eq(\ref{Eq19}) and $A$ and $B$ fitting parameters in analogy to our fits of galactic RC's per Eq(\ref{GRansatz}) (or Eq(\ref{GRansatz1}) for Figure \ref{Perseus}). This value of $M_p(r)$ is then compared to that of Eq(\ref{Eq20}). Of course, one can solve analytically for the exact correction $f(r)$ needed in the integral of Eq(\ref{Eq18}) to render Eq(\ref{Eq20}), i.e.,
\begin{equation}
f(r) = \frac{3\beta k T}{G \mu_A m_p 4 \pi \rho_o}\left(\frac{3}{r^2 + r_c^2}  - \frac{2 r^2}{\left(r^2 + r_c^2 \right)^2} \right) \left(1 + \left(\frac{r}{r_c}\right)^2 \right)^{1.5\beta} \label{exactclustercorrection}
\end{equation}
It's clear that our GR ansatz correction Eq(\ref{GRansatz}) in Eq(\ref{Eq23}) is nowhere near as complex as $f(r)$ in Eq(\ref{exactclustercorrection}) required in even this idealized situation. Thus, it is perhaps surprising that it works as well as it does (Table \ref{GRclusters}). 

\begin{figure}
\begin{tabular}{ll}
\includegraphics[width=3.25in]{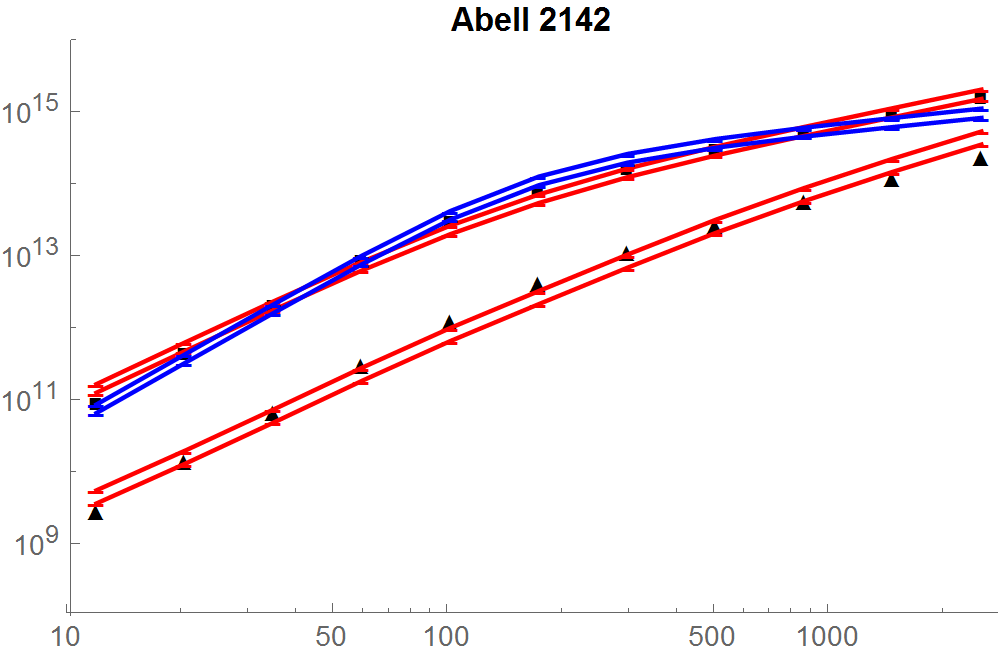} &
\includegraphics[width=3.25in]{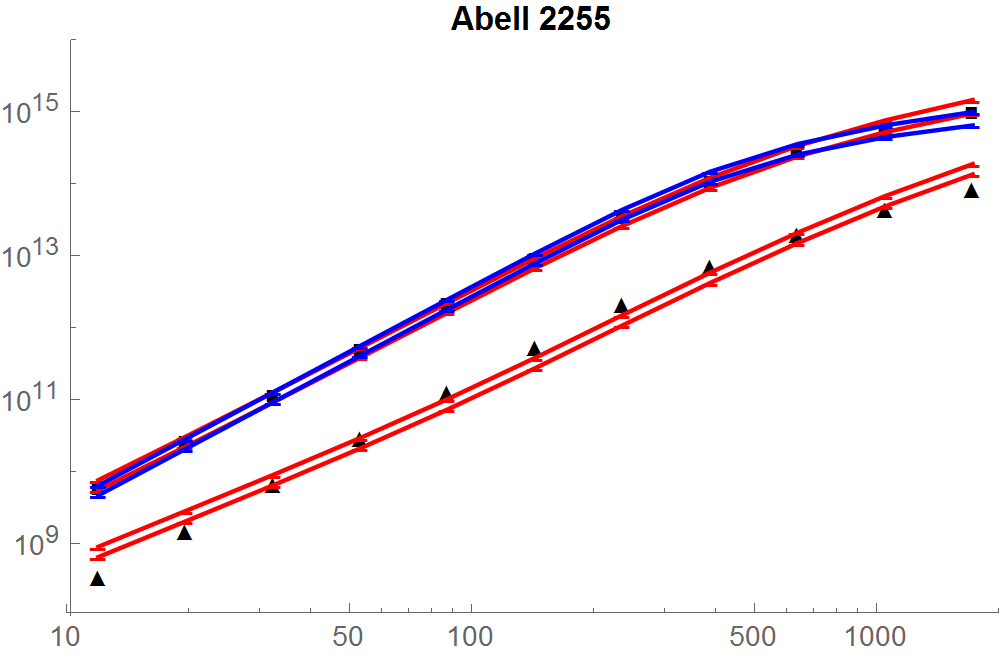} \\
\includegraphics[width=3.25in]{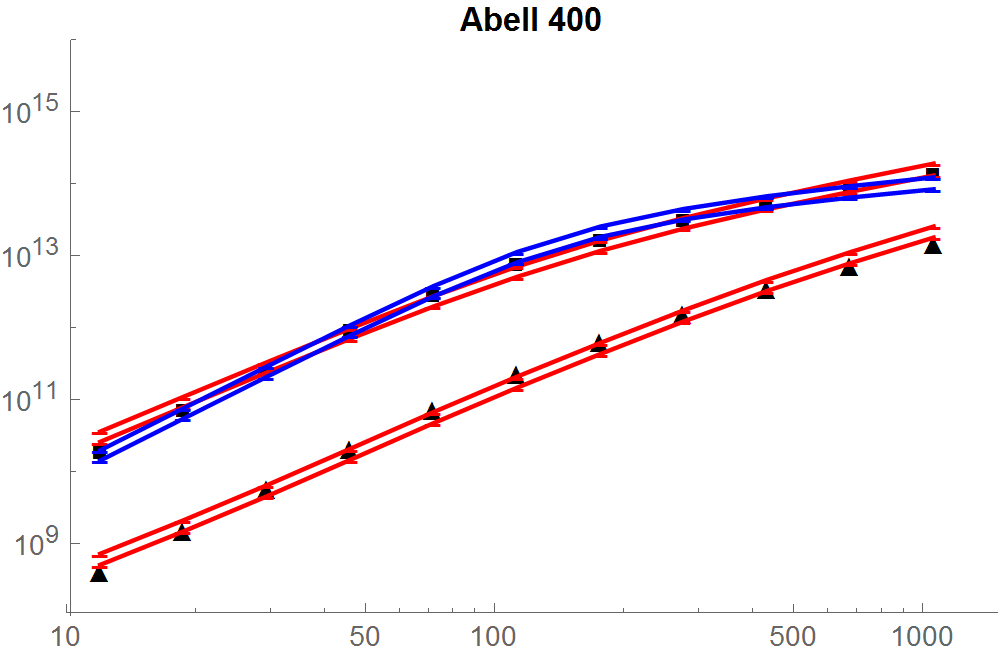} &
\includegraphics[width=3.25in]{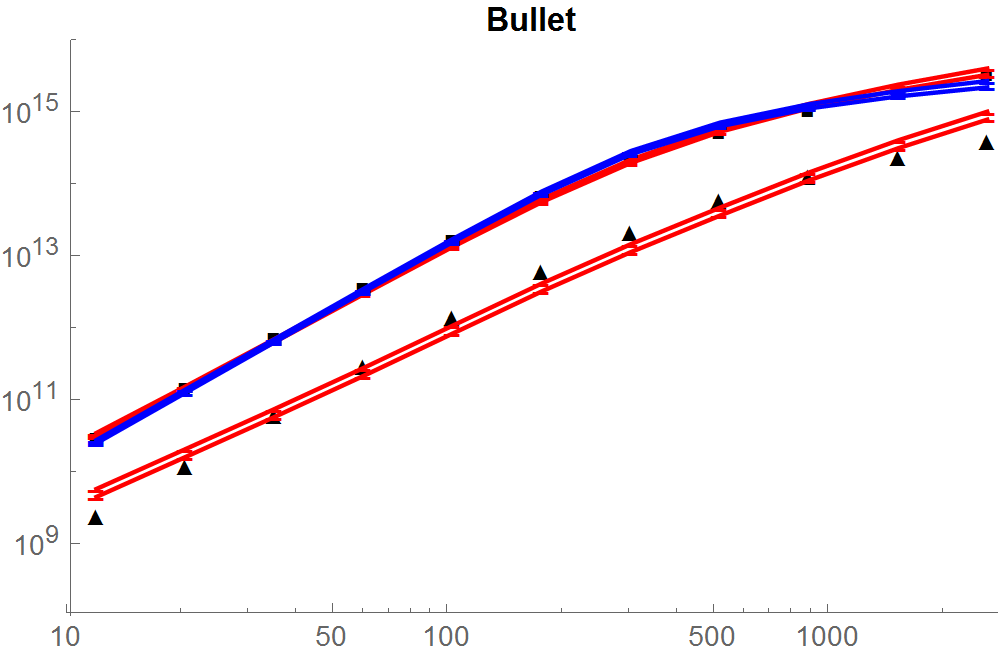} \\
\includegraphics[width=3.25in]{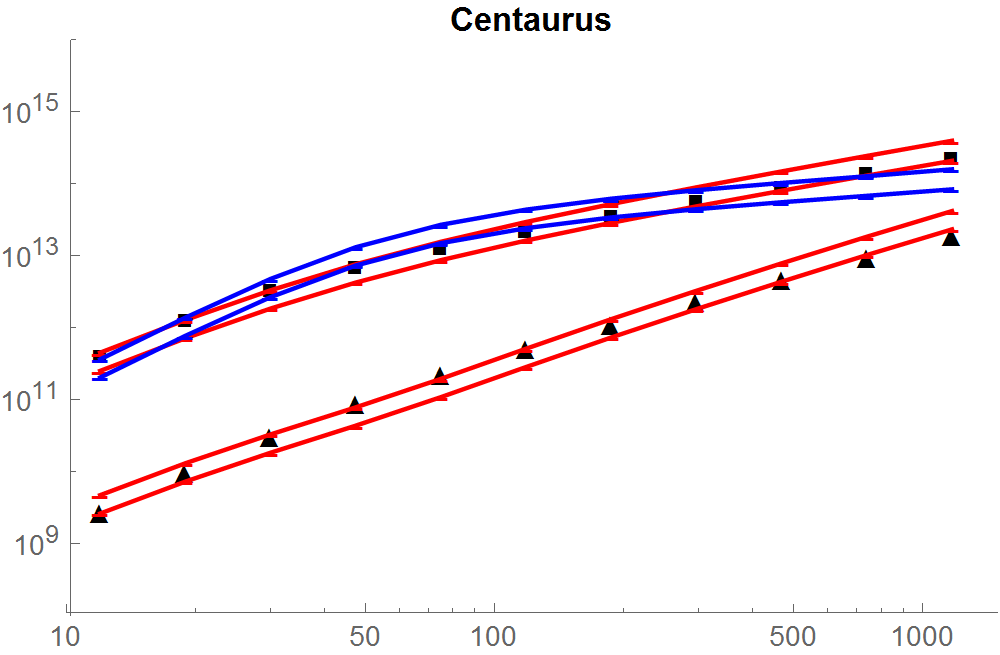} &
\includegraphics[width=3.25in]{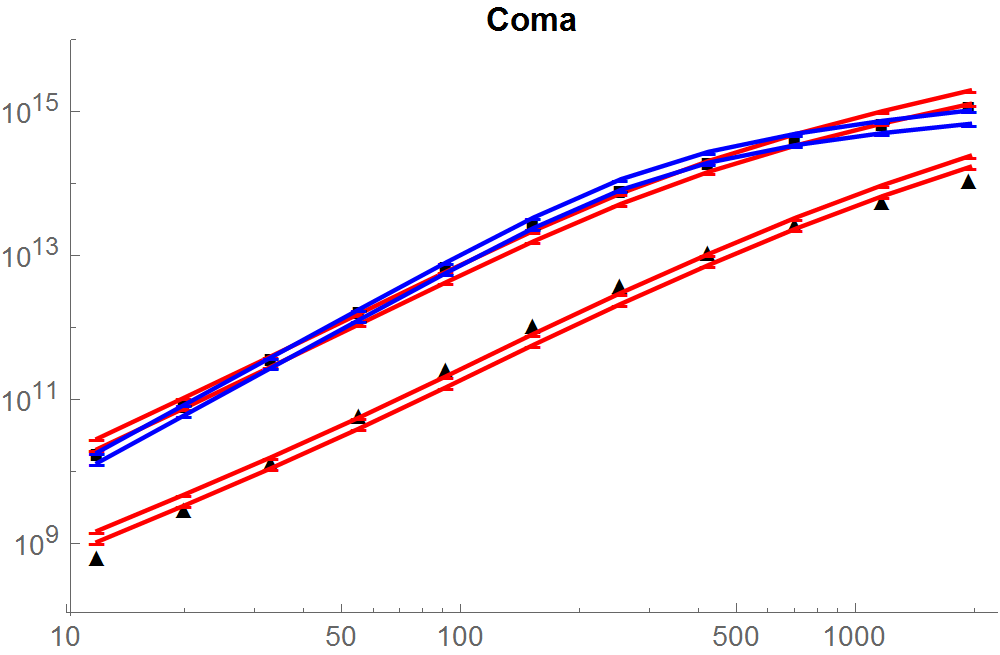}
\end{tabular}
\caption{Log-log plots of our GR ansatz fits, MSTG fits, and core-modified NFW DM fits of X-ray cluster mass profiles (compiled from ROSAT and ASCA data). Vertical scale is in solar masses and horizontal scale is in kpc. Our ansatz (upper red lines) is increasing the gas (dynamic) mass (triangles) to fit the proper mass (squares). MSTG (lower red lines) is decreasing the proper mass to fit the gas (dynamic) mass. Core-modified NFW DM (upper blue lines) is adding matter to increase the gas (dynamic) mass to fit the proper mass. The sizes of the objects are approximately equal to their errors. Line separation in the pair of lines (connecting fit points) corresponds to error. Corresponding numerical results are in Tables \ref{GRclusters}, \ref{mstgclusters}, \& \ref{DMclusters}.} \label{cluster1}
\end{figure}

Here we will compare (via tables and figures) our GR ansatz Eq(\ref{GRansatz}) to core-modified NFW and MSTG for fitting the mass profiles of the eleven X-ray clusters found in Brownstein \cite{brownstein} as obtained from Reiprich and B\"{o}hringer \cite{reiprich1, reiprich2} using combined ROSAT and ASCA data. All data were taken from Brownstein \cite{brownstein} wherein Brownstein found MSTG fits of these X-ray clusters bested those of MOND and STVG, while core-modified NFW bested MSTG. As shown in Tables \ref{mstgclusters} \& \ref{DMclusters}, we also found core-modified NFW bested MSTG. As with our comparison of our GR ansatzes to MOND, Burkett, and NFW for galactic RC's, one should keep in mind that MSTG, Burkett, and NFW are well-developed theories, so they are more constrained than our ansatz; the comparison between our ansatz, MSTG and core-modified NFW here is only for establishing the feasibility of our approach. In each case, we computed dynamic (or ``gas'') mass $M(r)$ and proper mass $M_p(r)$ at eleven logrithmically evenly spaced points from 13 kpc to $r_{out}$, ``the point where the density of the X-ray cluster drops to $\approx 10^{-28}$ g cm$^{-3}$ which is about 250 times the mean cosmological density of baryons.'' The values of $T$, $\rho_o$, $\beta$, and $r_c$, supplied by Brownstein, were used with Eqs(\ref{Eq18}-\ref{Eq20}) to compute $M_p(r)$ and $M(r)$. We then computed the MSTG best fit values of $M_o$ and $r_o$ reducing $M_p(r)$ to $M(r)$ at these eleven locations using Eqs(\ref{Eq21} \& \ref{Eq22}). Our best values of $A$ and $B$ were found by fitting the proper mass $M_p(r)$ of Eq(\ref{Eq23}). As with galactic RC's, we computed a simple mean square error (MSE) to quantify the fits. The plots of these best fits are in Figures \ref{cluster1} \& \ref{cluster2}. Our GR ansatz results are in Table \ref{GRclusters}, the MSTG results are in Table \ref{mstgclusters}, and the core-modified NFW fits are in Table \ref{DMclusters}. The results are discussed in section \ref{sec6} after obtaining fits of the CMB angular power spectrum (section \ref{sec5}).

\begin{table}
\begin{center}
\begin{tabular}{c|ccc}
Name & $A$ & $B$ & MSE \\
\hline
Bullet& 826 & 87.5 & 0.00221 \\
Abell 2142& 21500 & 34.0 & 0.00786 \\
Coma& 10500 & 198 & 0.0105 \\
Abell 2255& 1430 & 202 & 0.00256 \\
Perseus& 27800 & 0.00 & 0.00173 \\
Norma& 30300 & 105 & 0.0159 \\
Hydra-A& 30600 & 15.5 & 0.00081 \\
Centaurus& 130000 & 0.00 & 0.00737 \\
Abell 400& 47000 & 27.2 & 0.00848 \\
Fornax& 20900 & 2380 & 0.00126 \\
Messier 49& 999000 & 3440 & 0.00018 \\
\end{tabular}
\end{center}
\caption{GR ansatz Eq(\ref{GRansatz}) fits for mass profiles of X-ray clusters. $A$ is in kpc, $B$ is dimensionless and mean square error (MSE) is $(\Delta \mbox{Log}(M))^2$. Corresponding plots are in Figures \ref{cluster1} \& \ref{cluster2}.}
\label{GRclusters}
\end{table}

\begin{table}
\begin{center}
\begin{tabular}{c|ccc}
Name & $M_o$ & $r_o$ & MSE \\
\hline
Bullet& 69.15 (56.7) & 129.7 (116.8) & 0.0420 (0.0443) \\
Abell 2142& 29.64 (30.0) & 51.23 (56.8) & 0.0238 (0.0302) \\
Coma& 43.10 (30.7) & 97.86 (88.2) & 0.0294 (0.0342) \\
Abell 2255& 41.59 (43.8) & 154.6 (157.4) & 0.0385 (0.0386) \\
Perseus& 13.32 (10.7) & 24.05 (23.5) & 0.0192 (0.0239) \\
Norma& 24.01 (30.1) & 79.51 (97.6) & 0.0198 (0.0373) \\
Hydra-A& 9.84 (9.5) & 21.39 (23.9) & 0.0283 (0.0348) \\
Centaurus& 12.39 (10.0) & 14.86 (14.2) & 0.0166 (0.0218) \\
Abell 400& 6.63 (6.0) & 45.15 (44.7) & 0.0113 (0.0124) \\
Fornax& 9.91 (13.7) & 53.76 (67.4) & 0.0128 (0.0318) \\
Messier 49& 8.39 (10.3) & 8.55 (10.8) & 0.0176 (0.0244) \\
\end{tabular}
\end{center}
\caption{MSTG fits for mass profiles of X-ray clusters. $M_o$ is in $10^{14} M_{\odot}$, $r_o$ is in kpc and mean square error (MSE) is $(\Delta \mbox{Log}(M))^2$. $M_o$ and $r_o$ in parentheses resulting in MSE in parentheses were taken from Brownstein \cite{brownstein} minimizing reduced $\chi^2$. Corresponding plots are in Figures \ref{cluster1} \& \ref{cluster2}.}
\label{mstgclusters}
\end{table}

\begin{table}
\begin{center}
\begin{tabular}{c|ccc}
Name & $\rho_o$ & $r_s$ & MSE \\
\hline
Bullet& 3.32 (2.73) & 281 (328) & 0.00483 \\
Abell 2142& 10.1 (8.34) & 124 (142) & 0.0117 \\
Coma& 2.13 (1.91) & 238 (262) & 0.00397 \\
Abell 2255& 0.728 (0.67) & 375 (403) & 0.00167 \\
Perseus& 31.7 (28.8) & 60.8 (58.7) & 0.0219 \\
Norma& 1.73 (1.58) & 207 (223) & 0.00407 \\
Hydra-A& 32.8 (22.8) & 49.1 (59.3) & 0.0217 \\
Centaurus& 39.7 (38.0) & 39.2 (36.3) & 0.0237 \\
Abell 400& 2.32 (2.09) & 111 (120) & 0.00498 \\
Fornax& 1.52 (1.39) & 108 (117) & 0.00128 \\
Messier 49& 67.6 (36.8) & 15.3 (21.3) & 0.00749 \\
\end{tabular}
\end{center}
\caption{Core-modified NFW DM fits for mass profiles of X-ray clusters. $\rho_o$ is in $10^{6} M_{\odot}\:\mbox{kpc}^{-3}$, $r_s$ is in kpc and mean square error (MSE) is $(\Delta \mbox{Log}(M))^2$. Values in parentheses are from Brownstein's fits minimizing reduced $\chi^2$ \cite{brownstein}. Corresponding plots are in Figures \ref{cluster1} \& \ref{cluster2}.}
\label{DMclusters}
\end{table}

\begin{figure}
\begin{center}
\includegraphics [height = 60mm]{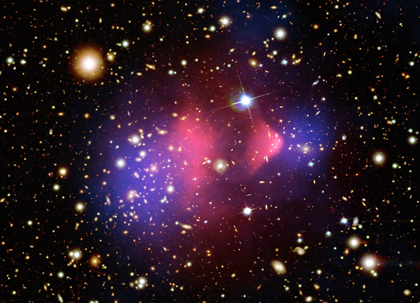}  \caption{Bullet Cluster X-ray and lensing composite (false color) image from NASA Release 06-297. Blue lobes are lensing data in the vicinity of the galaxies. Red lobes are X-ray images of ICM gas left behind after the bullet subcluster (right side) passed through the main cluster (left side).} \label{bullet}
\end{center}
\end{figure}

Before proceeding to the fits of the CMB angular power spectrum, we point out that gravitational lensing data of the Bullet Cluster (1E0657-558) originally touted as ``direct empirical proof of the existence of dark matter,'' \cite{clowe1} can be explained without DM \cite{moffat2, moffat8, brownstein2}. What happened in this case is a small galactic subcluster (``bullet cluster'') collided with a larger main cluster. The galaxies of both clusters passed through the collision region relatively unaffected, but the intracluster medium (ICM) gas of the two clusters was left behind in the collision region due to ram pressure. The result was four lobes of baryonic matter aligned as follows: the galaxies of the main cluster, the gas of the main cluster, the gas of the subcluster, and the galaxies of the subcluster (Figure \ref{bullet}). If one accepts that the mass of the cluster galaxies is only 10\% of the total baryonic mass, then in the absence of DM one would expect gravitational lensing maps of this region (blue lobes in Figure \ref{bullet}) to overlap X-ray images of the gas lobes (red lobes in Figure \ref{bullet}), since the gas possesses 90\% of the total baryonic mass. What Clowe et al. rather found \cite{clowe1} was that the lensing peaks were located in the galaxy lobes, so the galaxies are inside the blue lobes of Figure \ref{bullet}. Their conclusion was that there exists large quantities of DM which passed through the collision with the galaxies. Brownstein \&  Moffat \cite{moffat2, brownstein2} and Israel \& Moffat \cite{moffat8} explained the offset lensing peaks using MOG because $G(r)$ associated with the galaxies increases more than $G(r)$ associated with the gas, since the galaxies are farther removed from the center of the Newtonian gravitational potential. The explanation per our heuristic is that the ram pressure that swept the gas out of the galactic clusters did not displace the small spatial curvature originally swept into the clusters via gravitationally collapsing matter. Thus, the spatial curvature resides in the galactic lobes rather than the gas lobes and that produces more lensing about the galactic lobes than the gas lobes. This same effect is responsible for anisotropies in the angular power spectrum of the CMB, as we will explain in the following section \ref{sec5}. This does not mean that the ICM gas \textit{cannot} carry spatial curvature in such collision processes. Indeed, Jee et al. found a significant ``dark core'' in the X-ray gas region of a cluster collision that was not associated with any bright cluster galaxies \cite{jee}. And,  Bradac et al. used both weak and strong lensing to find a non-negligible mass concentration coincident with the main X-ray peak in the Bullet Cluster \cite{bradac}. 

Of course, this is just a heuristic, the bottom line in GR is that one must find a metric and stress-energy tensor that satisfy Einstein's equations everywhere on the spacetime manifold to have an allowable configuration. Without a complete spatiotemporal solution (allowable configuration), there is no way to tell a definitive corresponding dynamical story. In order to model gas having collapsed into dynamical collections of stars, galaxies, clusters, super clusters, filaments, etc., with differing values of mass as shown above, we would need multiply connected spacetime regions well beyond the simple FLRW-Schwarzschild combination. Unfortunately, there is likely no such exact solution forthcoming, so a GR-motivated ansatz may be the best we can do in exploring this proposition. In any event, a fit of the Bullet lensing offset using our GR ansatz at this point is of no interest, since the number of fitting parameters would equal the number of data points. Thus, fitting these lensing data will have to wait until the fitting parameters of our ansatz are constrained by theory or observations otherwise.

\section{CMB Angular Power Spectrum}\label{sec5}
\begin{figure}
\begin{center}
\includegraphics [height = 45mm]{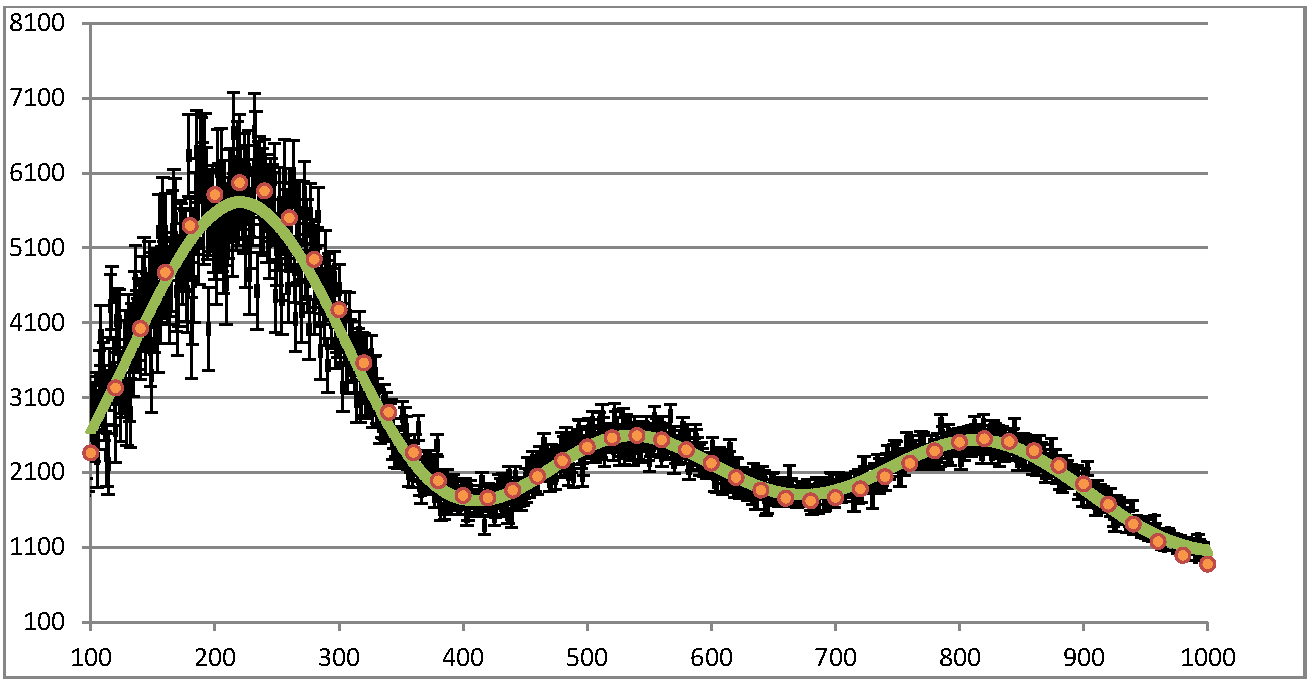}  \caption{This is a plot of $D_{\ell}$ in $\left(\mu K \right)^2$ versus $\ell$ in the range $100 \leq \ell \leq 1000$ for the Planck 2015 CMB data \cite{planckdata} (black error bars), the Planck consortium's best $\Lambda$CDM fit \cite{planckfit} (solid line), and our best HuS sCDM fit (dots). Our GR ansatz and STVG trivially reproduce the HuS fit without DM. Since we do not have $\Lambda$, our best fit to these data would equal the HuS sCDM best fit. The root-mean-square error (RMSE) for the HuS sCDM fit points shown is 225 $\left(\mu K \right)^2$. STVG can also trivially replace DM in $\Lambda$CDM and STVG keeps $\Lambda$, so the STVG best fit to these data would equal the $\Lambda$CDM best fit. The RMSE for the $\Lambda$CDM fit shown corresponding to the HuS fit points shown is 240 $\left(\mu K \right)^2$, although this fit is for all $\ell$ in the range $30 \leq \ell \leq 2508$.} \label{PlanckDataFit}
\end{center}
\end{figure}

After the Bullet Cluster, the most compelling evidence for DM is perhaps the CMB angular power spectrum \cite{carroll}. In this situation, small primordial inhomogeneities $\Delta_i = \frac{\delta \rho_i}{\rho_i}$ in the otherwise homogeneous energy density $\rho$ of the the background FLRW cosmology model lead to anisotropies in the CMB angular power spectrum $C_{\ell}$ versus $\ell$ (typically plotted $D_{\ell}$ versus $\ell$ where $D_{\ell} = \frac{\ell(\ell + 1) C_{\ell}}{2 \pi}$). The various components $\Delta_i$ typically considered are photons, baryons (this term includes electrons), neutrinos, and DM. Before recombination, the photons are coupled to the baryons and together their perturbations oscillate, gravity pulling them in while the photon gas pressure pushes them out. The neutrino perturbations simply stream free and the DM perturbations stay put, since DM is not affected by the photon gas pressure. As a result, the DM acts to enhance the contraction phase of the photon-baryon $\Delta_i$ oscillations while suppressing the rarefaction phase which manifests itself as enhanced odd peaks relative to even peaks in $D_{\ell}$ versus $\ell$. Thus, the size of the third peak relative to the second peak is taken to be an indication of DM and the Planck 2015 CMB data clearly show that the second and third peaks have very nearly the same amplitude (Figure \ref{PlanckDataFit}) in accord with the existence of DM. Therefore, it is this effect that we wish to reproduce without DM using our GR ansatz Eq(\ref{GRansatz}) and STVG.

The relativistic version of MOND known as the Tensor-Vector-Scalar (TeVeS) theory, does not show such an enhanced third peak \cite{skordis, skordis2}, so the Planck 2015 CMB data would appear to present a serious challenge for MOND \cite{mcgaugh}, although Skordis is working on a new approach and expects results in 2017 \cite{skordis3}. Angus \cite{angus} has argued that non-relativistic MOND would have no cosmological effect and one can obtain the enhanced third peak using a sterile neutrino with mass 11eV. This also allows MOND to account for ``the dark matter of galaxy clusters without influencing individual galaxies, ... and potentially fit the matter power spectrum.'' Whether this constitutes a true extension of MOND or a ``cheat'' is subject to debate, as cosmology certainly resides in the relativistic domain, so some relativistic version of MOND that accounts for the CMB anisotropies is needed \cite{mcgaugh2}.  

The situation with MOG is less contentious, as its practitioners seem to agree that it has addressed astronomical data at all scales with some degree of success. Recently, Moffat \& Toth \cite{toth} used STVG to successfully account for the matter power spectrum per data from the Sloan Digital Sky Survey, as well as the CMB angular power spectrum per data from WMAP and the Boomerang experiment for $\ell \leq$ 1000, which includes the third peak. In both cases, they used an enhanced gravitational constant $G$ in a fashion similar to the MOG account of the Bullet Cluster. Specifically, Moffat \& Toth \cite{toth} modified $G$ in Mukhanov's analytic approach \cite{mukhanov} such that, ``When a quantity containing $G$ appears in an equation describing a gravitational interaction, $G_{eff}$ must be used. However, when a quantity like $\Omega_b$ is used to describe a nongravitational effect, the Newtonian value of $G$ must be retained.'' The modification is given by
\begin{equation}
G_{eff} = G \left[1 + \alpha \left[1 - \left(1 + \frac{\mu a}{k} \right) \exp{\left(-\frac{\mu a}{k}\right)} \right] \right] \label{Geff}
\end{equation}
$\Omega_o$ is the current total energy density divided by the critical energy density (required to make flat spatial hypersurfaces in FLRW cosmology). $\Omega_m$ is the current contribution to the total energy density due to matter. In the model herein, we have $\Omega_m = \Omega_o$, i.e., $\Omega_{\Lambda} = 0$ and the current photon and neutrino contributions are negligible. $\Omega_b$ is the current contribution to the total energy density due to baryons, so for our ansatz and STVG, $\Omega_b = \Omega_m$, i.e., $\Omega_{DM} = 0$. Since STVG is the best modified gravity fit of the CMB angular power spectrum to date, we will compare our fitting procedure to that of STVG. As it turns out, both can reproduce the standard cold dark matter (sCDM) fit of the first three acoustic peaks in a rather trivial fashion. To show this, we need to look carefully at an sCDM fitting procedure. Before we start, we point out that STVG gets rid of DM but it retains $\Lambda$ while we get rid of both DM and $\Lambda$. Therefore, the comparison of STVG and our ansatz in this context will only involve the replacement of DM in sCDM. 

The analytic method of Hu \& Sugiyama \cite{husugiyama, husugiyama2} (HuS) maps closely to the full numerical solution in the first three acoustic peaks and its physics is transparent, so we will use HuS. Since the Planck 2015 data strongly suggest a spatially flat universe \cite{planck}, i.e., $\Omega_o = 1$, we will work in that context. Further, we have no $\Lambda$, so $\Omega_m = \Omega_o = 1$ while a proper STVG fit would have $\Omega_m \approx 0.3$ and $\Omega_{\Lambda} \approx 0.7$. Here is a brief outline of HuS in this context with an explanation of our GR ansatz and STVG modifications thereto.

The spatially flat FLRW metric (we will use c = 1 in what follows) \cite{huphd}
\begin{equation}
ds^2 = -dt^{2} + a^2(t)\delta_{ij} dx^idx^j\label{frwmetric}
\end{equation}
is modified in the conformal Newtonian gauge by the existence of the perturbations to read
\begin{equation}
ds^2 = -\left(1 + 2 \Psi \right)dt^{2} + \left(1 + 2 \Phi \right)a^2(t)\delta_{ij} dx^idx^j \label{cmbmetric}
\end{equation}
where $\Psi$ is called ``the Newtonian potential'' and $\Phi$ is a perturbation to the spatial curvature. When pressure is negligible, $\Psi = -\Phi$. Since the spatial part of the Schwarzschild metric can be written \cite{wong} 
\begin{equation}
ds^2 = \left(1 + \frac{GM}{2r}\right)^4\delta_{ij} dx^idx^j \label{schwarzschild2}
\end{equation}
we have for small $\displaystyle \frac {GM}{2r} \left(= -\frac{\Phi}{2} \right)$  
\begin{equation}
ds^2 \approx \left(1 + \frac{2GM}{r}\right)\delta_{ij} dx^idx^j \label{schwarzschild3}
\end{equation}
which explains the perturbation $\Phi$ in Eq(\ref{cmbmetric}) as a small Newtonian potential placed in a spatially flat FLRW background. Thus, as with galactic rotation curves and X-ray cluster mass profiles, this missing mass phenomenon constitutes a small geometric effect, i.e., $\Psi$ and $\Phi$. STVG will be correcting $G$ as shown above. Einstein's equations directly relate $\Psi$ and $\Phi$ to the total energy density $\rho$, which only appears coupled to $G$. That is, Einstein's equations give
\begin{equation}
k^2 \Phi = 4 \pi G \rho \left(\frac{a}{a_o}\right)^2 \Delta_T \label{EE1}
\end{equation}
and
\begin{equation}
\Phi + \Psi = -\frac{8 \pi G \rho}{k^2} \left(\frac{a}{a_o}\right)^2 \Pi \label{EE2}
\end{equation}
in the total matter rest frame gauge in $k$ space where $\Pi$ is the anisotropic stress and $\Delta_T$ is the total perturbation. Thus, a multiplicative correction to $\Psi$ and $\Phi$ on the LHS of Eqs(\ref{EE1} \& \ref{EE2}) per our ansatz could equally be viewed as a multiplicative STVG correction to $G$ on the RHS. Eqs(\ref{EE1} \& \ref{EE2}) are represented in HuS as
\begin{equation}
\bar{\Phi}(a,k) = \frac{9}{10} \left(\frac{k_{eq}}{k}\right)^{3/2} \frac{a + 1}{a^2} \left[1 + \frac{2}{5} f_{\nu} \left(1 - \frac{1}{3} \frac{a}{a + 1}\right)\right] U_G(a) \label{Phibar}
\end{equation}
and 
\begin{equation}
\bar{\Psi}(a,k) = -\frac{6}{5} \left(\frac{k_{eq}}{k}\right)^{2} \frac{a + 1}{a^2} f_{\nu} \frac{\bar{N}_2}{(a + 1)} - \bar{\Phi}(a,k) \label{Psibar}
\end{equation}
where $f_{\nu}$ is the fraction of the radiation energy density contributed by the neutrinos (0.405),
\begin{equation}
U_G(a) =\frac{9a^3+2a^2-8a-16+16\sqrt{a+1}}{9a(a+1)} 
\end{equation}
and
\begin{equation}
\bar{N}_2 = \frac{12}{45a}\frac{\sqrt{k}}{\sqrt{\text{keq}}} \left[8+4 a-3 a^2-8 \sqrt{1+a}+ 4 a \log \frac a4 +4 a \log \left(\frac{\sqrt{1+a}+1}{\sqrt{1+a}-1}\right)\right]
\end{equation}
is the neutrino quadrupole contribution to the anisotropic stess with 
\begin{equation}
k_{eq} = 7.46 \times 10^{-2} \left(\frac{\text{T}_o}{2.7}\right)^{-2} h^2 \Omega_o \label{keq}
\end{equation}
$T_o$ is the current temperature of the CMB and $h$ is the current Hubble constant divided by 100 km/s/Mpc. Again, Planck data strongly suggest $\Omega_o = 1$, so we will use that model. The overbar represents values prior to contributions from the matter transfer function $T(k)$. To account for that 
\begin{equation}
\Phi(a,k) =\bar{\Phi}\left((1-T(k))e^{-\alpha_1 \left(\frac{a k}{k_{eq}}\right)^{\beta}}+T(k)\right)
\end{equation}
and
\begin{equation}
\Psi(a,k) =\bar{\Psi}\left((1-T(k))e^{-\alpha_2 \left(\frac{a k}{k_{eq}}\right)^{\beta}}+T(k)\right)
\end{equation}
where
\begin{equation}
T(k) = \frac{\log (1+2.34 q)}{2.34q} \left [(6.71 q)^4+(5.46 q)^3+ (14.1 q)^2+3.89 q+1 \right ]^{-1/4} \label{Tk}
\end{equation}
with $\displaystyle q = k e^{2 \Omega_b}/(h^2 \Omega_m)$. $\alpha_1$, $\alpha_2$, and $\beta$ are fitting factors which Hu \& Sugiyama found to be 0.11, 0.097 and 1.6, respectively. Now we can find the monopole and dipole CMB temperature perturbations at recombination ($\Theta_0(\eta_{\star})$ and $\Theta_1(\eta_{\star})$, respectively) before damping by solving
\begin{equation}
\Theta_0(\eta) = -\frac{k}{3}\Theta_1(\eta) - \dot{\Phi}(\eta)
\end{equation}
and
\begin{equation}
\Theta_1(\eta) = -\frac{\dot{R}(\eta)}{1 + R(\eta)}\Theta_1(\eta)+\frac{1}{1 + R(\eta)}k \Theta_0(\eta) + k \Psi(\eta)
\end{equation}
for $k > 0.08 h^3$ and the same equations with $R(\eta) = 0$ for $k \leq 0.08 h^3$ from $\eta = 0$ to $\eta = \eta_{\star}$, which is the conformal time at recombination. [The derivatives are with respect to conformal time $\eta$.] We have
\begin{equation}
a(\eta) = \frac{1}{8} \left( 4\sqrt{2} k_{eq}\eta + k_{eq}^2 \eta^2 \right)
\end{equation}
and
\begin{equation}
R(\eta) = \frac{1}{1 - f_{\nu}} \frac{3\Omega_b}{4\Omega_o} a(\eta)
\end{equation}
where $a(\eta)$ is normalized at matter-radiation equality $a_{eq} = 1$. We also have
\begin{equation}
a(\eta_o) = \frac{\Omega_o h^2}{4 \times 10^{-5} \left(\frac{T_o}{2.7}\right)^{4}}
\end{equation}
and the redshift at recombination
\begin{equation}
z_{\star} = 1000 \Omega_b^{-c1/c2}
\end{equation}
with $c1 = 0.43$ and $c2 = 16 + 1.8 \text{log}(\Omega_b)$. The CMB temperature perturbations are then damped due to diffusion during recombination 
\begin{equation}
\Theta_0(\eta_{\star}) + \Psi(\eta_{\star}) \rightarrow e^{-\left(\frac{k}{k_D}\right)^m}\left(\Theta_0(\eta_{\star}) + \Psi(\eta_{\star})\right)
\end{equation}
\begin{equation}
\Theta_1(\eta_{\star}) \rightarrow e^{-\left(\frac{k}{k_D}\right)^m}\Theta_1(\eta_{\star})
\end{equation}
where
\begin{equation}
m = 1.46 \left(h^2 \Omega_o \right)^{0.0303} \left(1 - 0.128 \tan ^{-1}\left[\log \left(32.8 h^2 \Omega_b \right)^{0.643}\right]\right)
\end{equation}
\begin{equation}
k_D = F_1 \left(\frac{2}{\pi} \tan ^{-1}\left [\frac{\pi}{2} \left(h^2 \Omega_b \right)^{p_2} \left(\frac{F_2}{F_1}\right)^{p_2/p_1}\right] \right)^{p_1/p_2}
\end{equation}
$$
p_1 = 0.25, \qquad p_2 =2.38 \left(h^2 \Omega_o \right)^{0.184}
$$ 
\begin{equation}
F_1 = 0.293 \left(h^2 \Omega_o \right)^{0.545} \left(\frac{1}{\left(25.1 h^2 \Omega_o \right)^{0.648}}+1\right)
\end{equation}
and
\begin{equation}
F_2 = 0.524 \left(h^2 \Omega_o \right)^{0.505} \left(\frac{1}{\left(10.5 h^2 \Omega_o \right)^{0.564}}+1\right)
\end{equation}
Next the damped monopole and dipole perturbations at $\eta_{\star}$ are free streamed to higher $\ell$ today
\begin{multline}
\Theta_{\ell}(\eta_o) = \left[\Theta_o(\eta_{\star}) + \Psi(\eta_{\star})\right](2\ell + 1)j_{\ell}(k \Delta \eta_{\star}) + \\ \Theta_1(\eta_{\star}) \left[\ell j_{\ell - 1}(k \Delta \eta_{\star}) -  (\ell + 1) j_{\ell + 1}(k \Delta \eta_{\star})\right] + (2 \ell + 1) \text{ISW}
\end{multline}
where $j_{\ell}(z)$ is the spherical Bessel function, $\Delta \eta_{\star} = \eta_o - \eta_{\star}$, and the integrated Sachs-Wolfe effect is
\begin{equation}
\text{ISW} = \int_{\eta_{\star}}^{\eta_o} \left[\dot{\Psi} - \dot{\Phi}\right] j_{\ell}(k \Delta \eta) d\eta
\end{equation}
with $\Delta \eta = \eta_o - \eta$. We used the HuS approximation
\begin{equation}
\text{ISW} \approx \left[\Psi - \Phi \right]_{\eta_{\star}}^{\eta_o} j_{\ell}(k \Delta \eta_{\star})
\end{equation}
Finally, the $\Theta_{\ell}(\eta_o,k)$ are integrated to give 
\begin{equation}
\frac{2\ell + 1}{4\pi} C_{\ell} = \frac{V}{2 \pi^2} \int \frac{dk}{k} k^3 \frac{\left| \Theta_{\ell}(\eta_o,k) \right|^2}{2\ell + 1}
\end{equation}
where $V$ is an overall normalization constant. We used 1000 values of $k$ in the interval $\frac{\pi}{\eta_o} \leq k \leq k_D$ for each of the 46 points in the HuS fit shown in Figure \ref{PlanckDataFit}. Primarily, one has only $\Omega_b$ and $h$ for this HuS sCDM fit, but we adjusted a few others, viz., $m$ in the damping factor ($1.46 \rightarrow 1.10(5)$), the equation for redshift at decoupling ($1000 \rightarrow 1080(5)$), and $p1$ in the equation for the damping factor ($0.25 \rightarrow 0.23(1)$), to obtain our HuS fit. While $h$ is certainly smaller than accepted (0.49(1) instead of $\sim 0.7$) and $\Omega_b$ is too large (0.090(1) instead of $\sim 0.05$), the combination $h^2 \Omega_b$ is close to the accepted range ($0.02222 \pm 0.00023$) \cite{planck}. Again, the point of the fit isn't to find the fit parameters per se, although that will certainly be important once we are at the stage of building a cosmology model. Rather, we want to find out what change is required to bring the third peak up to the second peak without DM in the context of our GR ansatz and compare that with an established program in modified gravity, STVG, as well as the concordance model, $\Lambda$CDM. 

In order to remove the DM from this sCDM fit, we look to the very first step Eqs(\ref{Phibar} \& \ref{Psibar}) since that is the representation of Einstein's equations in the metric perturbations $\Psi$ and $\Phi$. Therein, we see that the only place matter of any kind enters the equations is through $k_{eq}$ by way of $h^2 \Omega_m$. Of course, that's because $h^2 \Omega_m \propto G \rho_m(a_o)$. Thus, eliminating DM from $\rho_m(a_o)$ means a loss of 91\% of the critical density necessary for $\Omega_m = \Omega_o = 1$ (in this particular HuS fit). The correction needed to restore the HuS fit, having removed the DM, is straightforward and exact. Since we have $\Omega_m \rightarrow \Omega_b$ in removing DM, in order to restore the HuS fit, we must multiply $h^2 \Omega_m$ by $\Omega_b^{-1}$, i.e., $\Omega_b^{-1}(h^2 \Omega_m \rightarrow h^2 \Omega_b) = h^2 = h^2 \Omega_m$. In order to have that change to $k_{eq}$ in our view, $\Phi$ and $\Psi$ are multiplied by $\Omega_b^{-3/2}$, since they are both proportional to $k_{eq}^{3/2}$. In STVG, we simply need $G \rightarrow \Omega_b^{-1}G$ which Eq(\ref{Geff}) gives for $\mu = 0$ and $\alpha = \Omega_b^{-1} - 1$ \footnote{The $k$ dependence is necessary for the matter power spectrum, so nonzero $\mu$ is required there.}. 

While our GR ansatz is couched in real space and this correction is in $k$ space, the correction factor is a constant, so it is the same in both spaces. Thus, our ansatz correction to $\Phi$ and $\Psi$ is simply $A = 0$ and $B = \Omega_b^{-3}$. As with the corrected Newtonian potentials for galactic RC's and X-ray cluster mass profiles, even though this correction is much larger than 1, the corrected $\Phi$ and $\Psi$ are still much less than 1 (they were already small including DM). Thus, this particular fit says that the mass of the baryons in the context of Big Bang nucleosynthesis is only approximately 10\% of that needed for critical density $\Omega_m = 1$, but the mass of that same matter in the FLRW context is precisely $\Omega_m = 1$. According to the heuristic, the density perturbations swept up spatial curvature during their formation. The photon pressure, apparently, leaves this spatial curvature behind as it pushes the baryons outward in rarefaction phases, just as ram pressure kept the ICM gas in the Bullet Cluster collision zone while the galaxies and spatial curvature passed through. As for why space is flat on cosmological scales, following the heuristic, it could be that most of the matter has already coalesced into stars, galaxies and clusters, thereby having swept up and localized the original spatial curvature leaving vast spatially flat regions in its wake. Exploring this possibility will require fits of the matter power spectrum and a complete cosmology model in future work.

Again, in STVG, wherever you have $\Omega_m h^2$ (i.e., you have $G \rho_m(a_o)$) in your fitting procedure, $G \rightarrow \Omega_b^{-1}G$ exactly restores the loss of $\Omega_{DM}$ from $\Omega_m$. In the standard best fit for Planck 2015, i.e., $\Lambda$CDM, $\Omega_m \approx 0.3$ and $\Omega_b  \approx 0.05$, so $G \rightarrow 6G$ allows baryons alone to do the work of DM plus baryons in the $\Lambda$CDM fit. That is what Moffat \& Toth found \cite{toth} using $\mu^{-1} \approx \text{particle horizon}$, which is $\mu \approx 0$, so $G \rightarrow (1 + \alpha)G$. Since STVG keeps $\Lambda$ (in fact it entails $\Lambda$, \cite{toth}), the STVG fit is equal to the $\Lambda$CDM fit in Figure \ref{PlanckDataFit}. We now discuss the results of these fits.

\section{Discussion}\label{sec6}
\begin{figure}
\begin{tabular}{ll}
\includegraphics[width=3.25in]{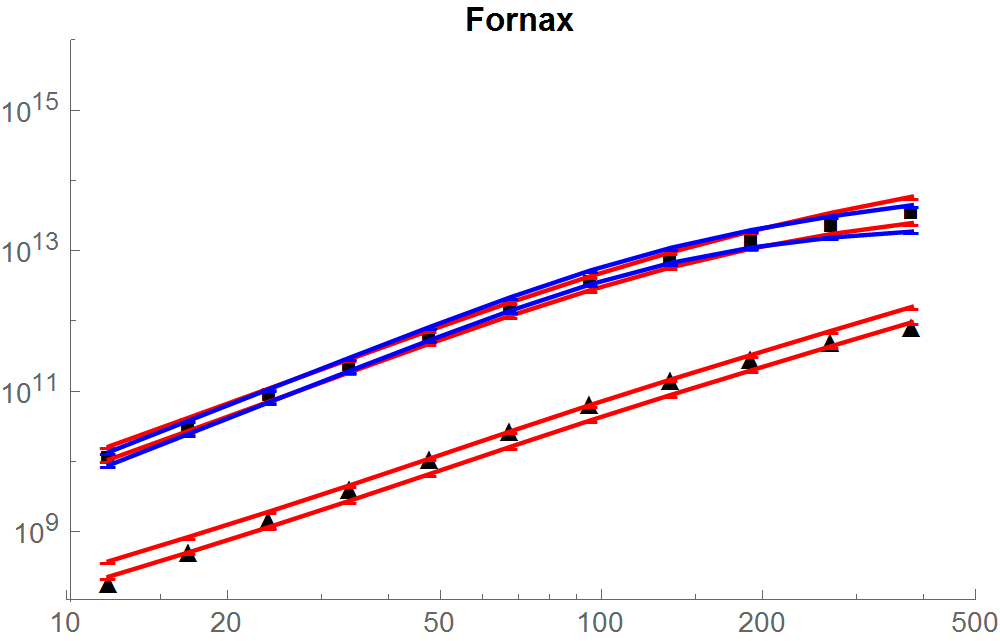} &
\includegraphics[width=3.25in]{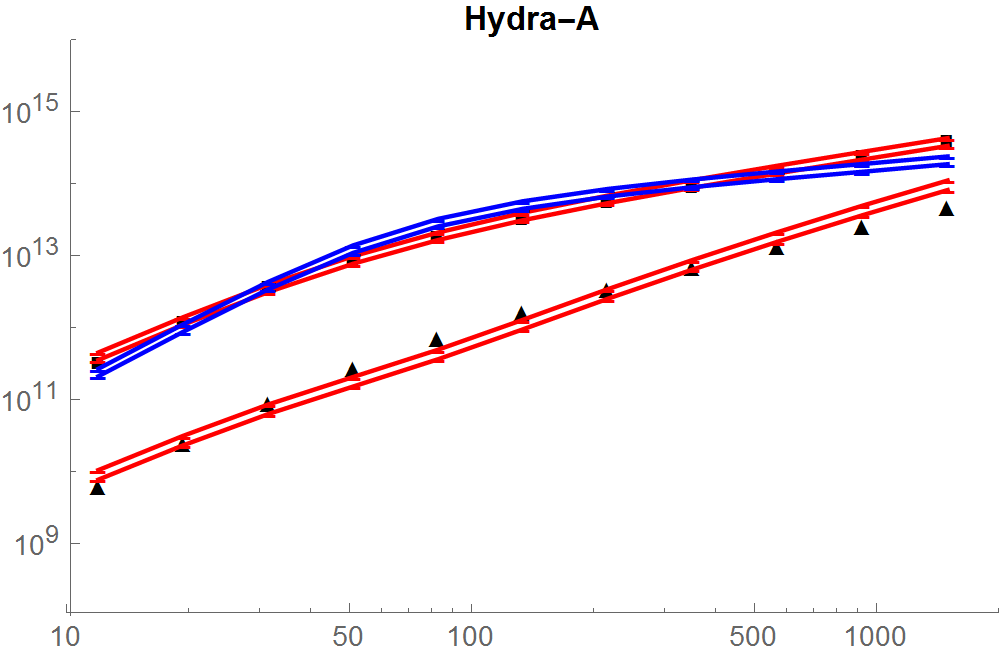} \\
\includegraphics[width=3.25in]{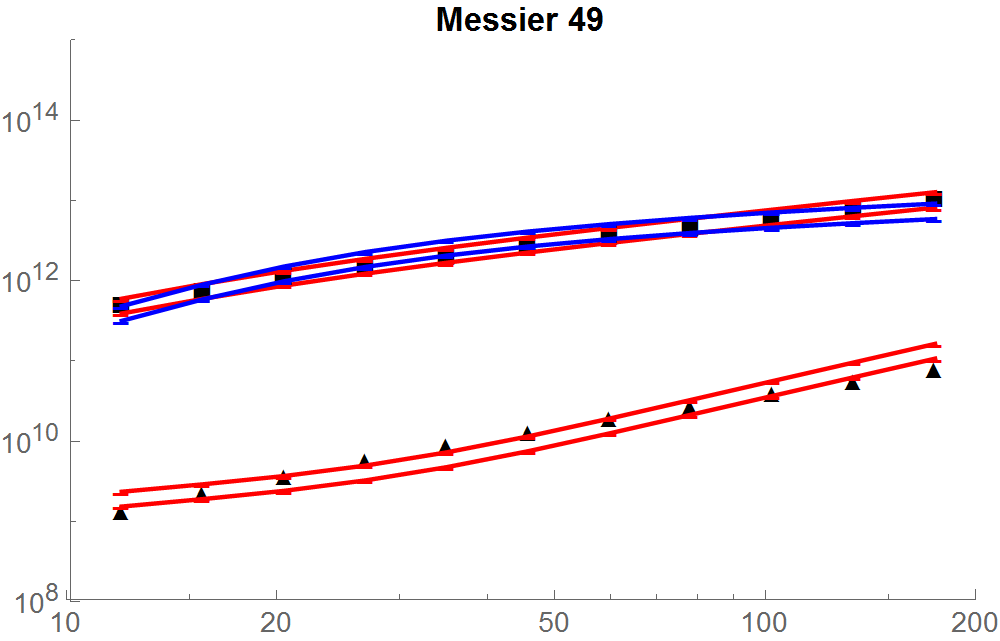} &
\includegraphics[width=3.25in]{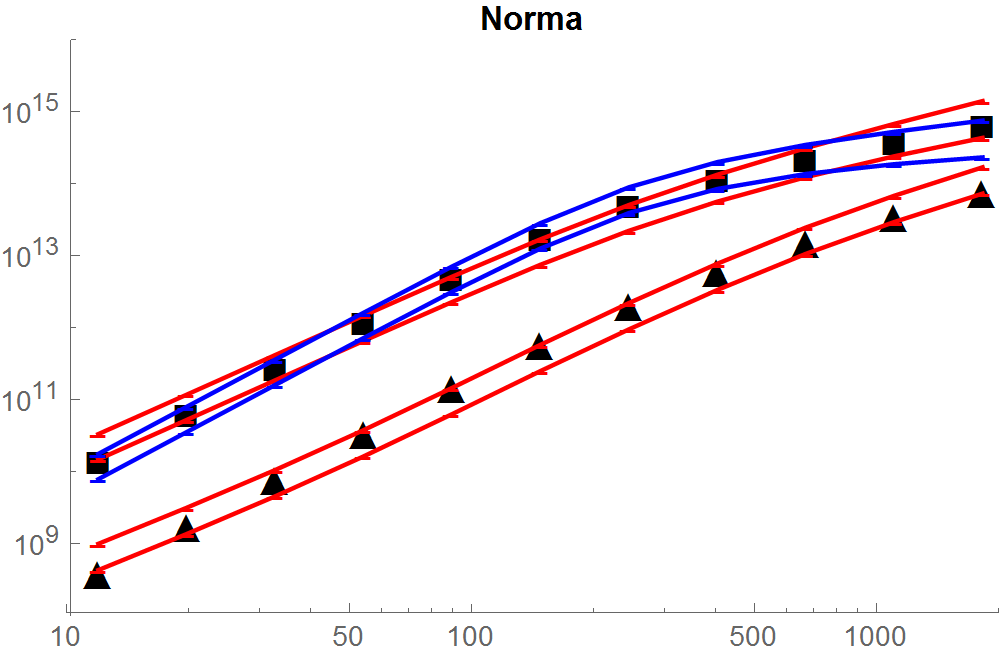} \\
\includegraphics[width=3.25in]{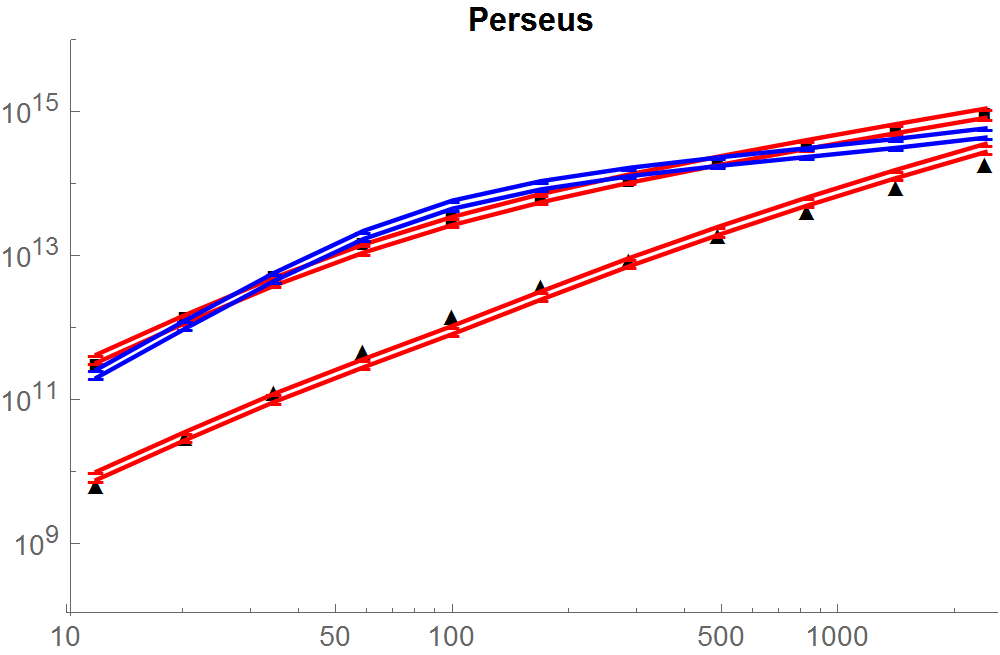} &
\end{tabular}
\caption{More log-log plots of our GR ansatz fits, MSTG fits, and core-modified NFW DM fits of X-ray cluster mass profiles (compiled from ROSAT and ASCA data). Vertical scale is in solar masses and horizontal scale is in kpc. Our ansatz (upper red lines) is increasing the gas (dynamic) mass (triangles) to fit the proper mass (squares). MSTG (lower red lines) is decreasing the proper mass to fit the gas (dynamic) mass. Core-modified NFW DM (upper blue lines) is adding matter to increase the gas (dynamic) mass to fit the proper mass. The sizes of the objects are approximately equal to their errors. Line separation in the pair of lines (connecting fit points) corresponds to error. Corresponding numerical results are in Tables \ref{GRclusters}, \ref{mstgclusters}, \& \ref{DMclusters}.} \label{cluster2}
\end{figure}

\begin{figure}
\begin{center}
\includegraphics[height = 50mm]{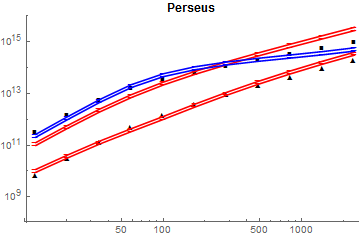} \caption{Log-log plot of the GR ansatz Eq(\ref{GRansatz1}) fit, MSTG fit, and core-modified NFW DM fit of the Perseus X-ray cluster mass profile. The MSTG and core-modified NFW DM fits are those seen in Perseus of Figure \ref{cluster2}. The Eq(\ref{GRansatz1}) fit produced an MSE of 0.114 compared to the GR ansatz Eq(\ref{GRansatz}) fit for Perseus in Figure \ref{cluster2} of 0.00173.} \label{Perseus}
\end{center}
\end{figure}

Overall, the GR ansatz Eq(\ref{GRansatz}) fits of THINGS data (average MSE = 101 $(\mbox{km/s})^2$, Table \ref{GRRC}) compare well with MOND (d unconstrained average MSE = 51.7 $(\mbox{km/s})^2$ and d constrained average MSE = 67.1 $(\mbox{km/s})^2$, Table \ref{mondRC}), Burkett DM halo (119 $(\mbox{km/s})^2$, Table \ref{BurkettRC}), and NFW DM halo (149 $(\mbox{km/s})^2$, Table \ref{NFWRC}). The $A_i$ for our GR ansatz fits of THINGS data (Table \ref{GRRC}) are not surprising. It is well known that the bulge has little effect on the missing mass problem and $A_{bulge}$ is zero when bulge data are available. Karukes \& Salucci ignored the bulge contribution altogether \cite{karukes}. Of course, since that is the case and our GR ansatz Eq(\ref{GRansatz}) actually gets larger for smaller $r$, it is necessary that $A_{bulge}$ = 0 in this case. The results for GR ansatz Eq(\ref{GRansatz1}), which gets smaller for smaller $r$, contain some nonzero $A_{bulge}$ values (Table \ref{GRRC1}) and this GR ansatz produced a slightly better average MSE of 81.9 $(\mbox{km/s})^2$ versus 101 $(\mbox{km/s})^2$ for Eq(\ref{GRansatz}). However, as is evident for the Perseus cluster in Figures \ref{cluster2} \& \ref{Perseus}, Eq(\ref{GRansatz1}) does not work well for X-ray cluster mass profiles while Eq(\ref{GRansatz}) compares well to alternatives in both cases. So, if we're looking for a single functional form for the GR ansatz, Eq(\ref{GRansatz1}) is ruled out. However, there is no reason a priori to assume the GR ansatz must have the same functional form for all scales.

\begin{figure}
\begin{tabular}{ll}
\includegraphics[width=3.5in,height=2.5in]{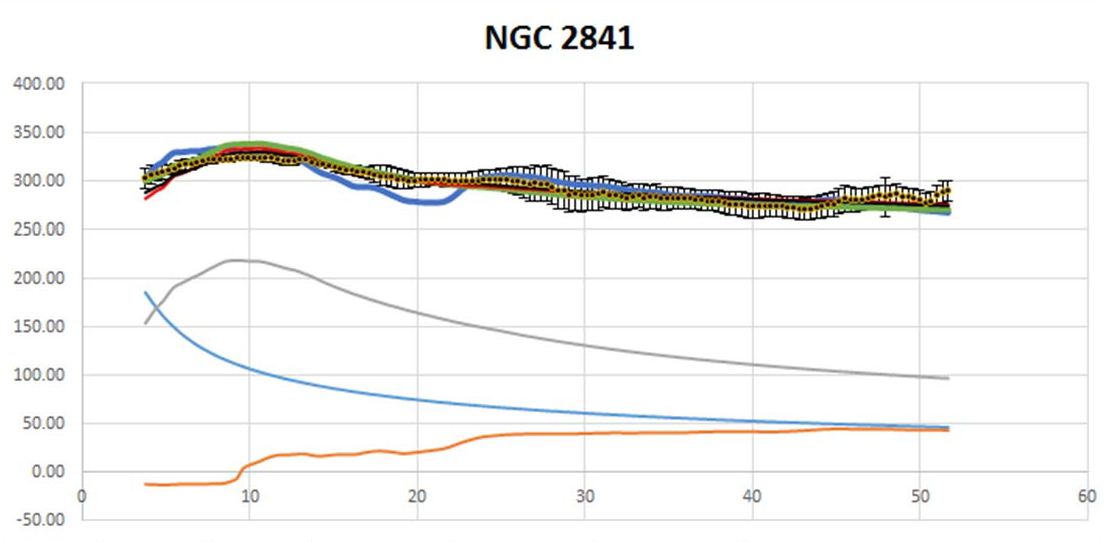} &
\includegraphics[width=3.5in,height=2.5in]{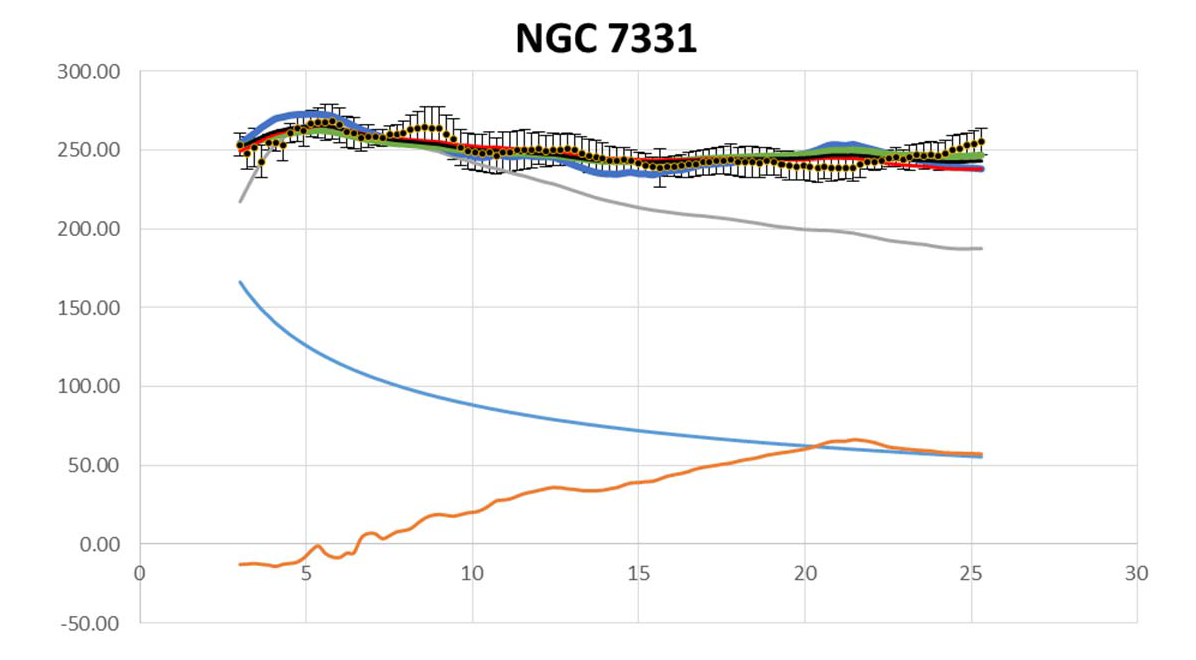} \\
\includegraphics[width=3.5in,height=2.5in]{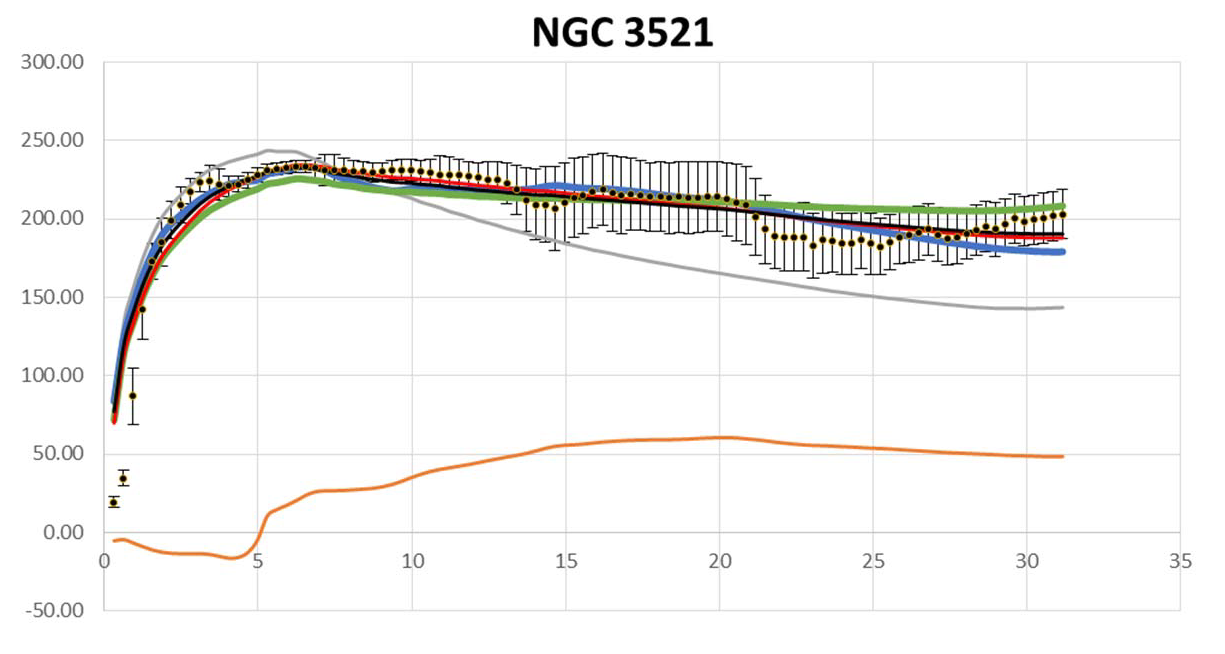} &
\includegraphics[width=3.5in,height=2.5in]{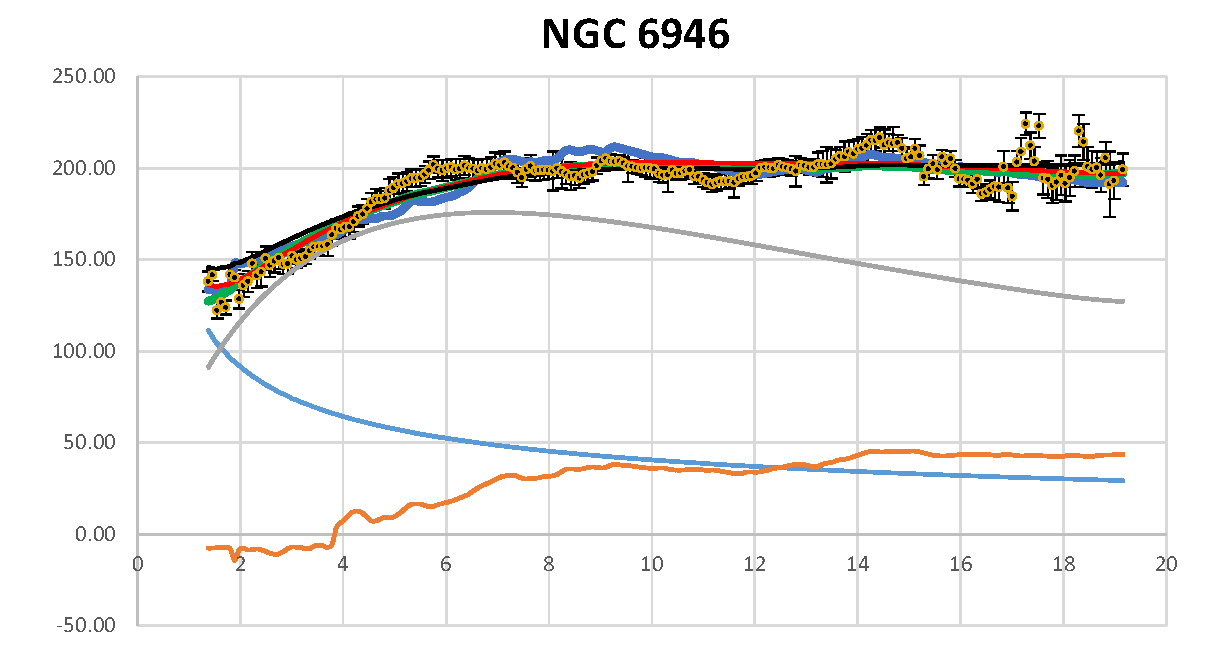} \\
\includegraphics[width=3.5in,height=2.5in]{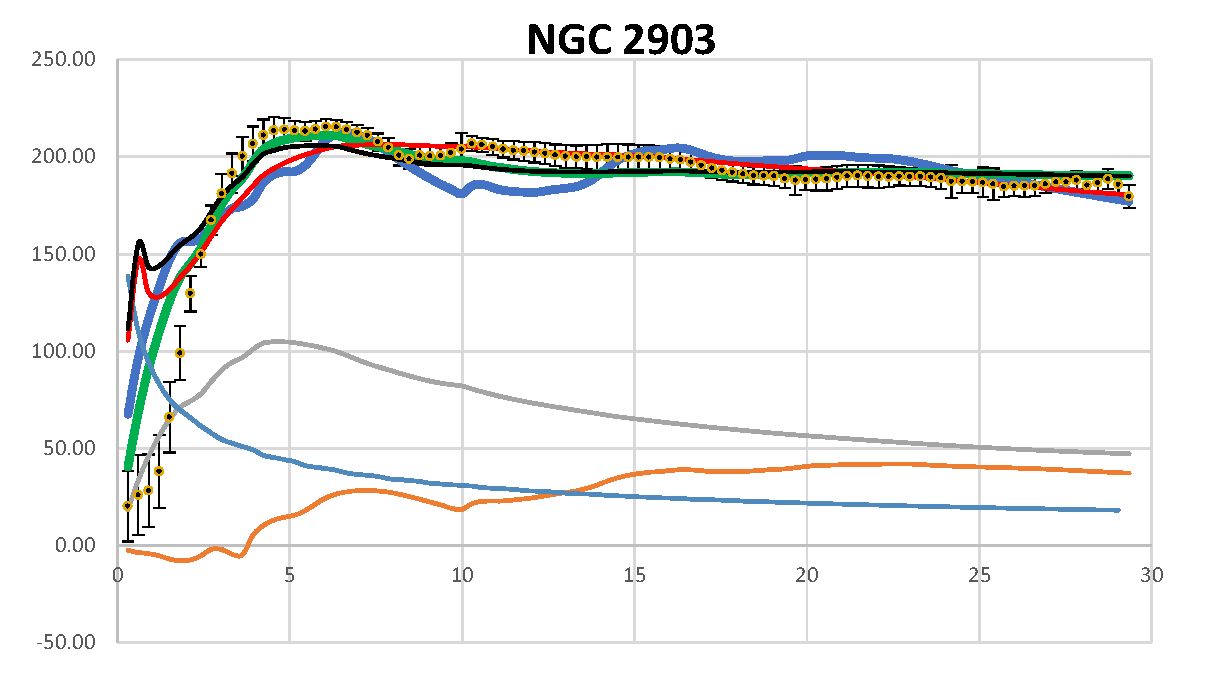} &
\includegraphics[width=3.5in,height=2.5in]{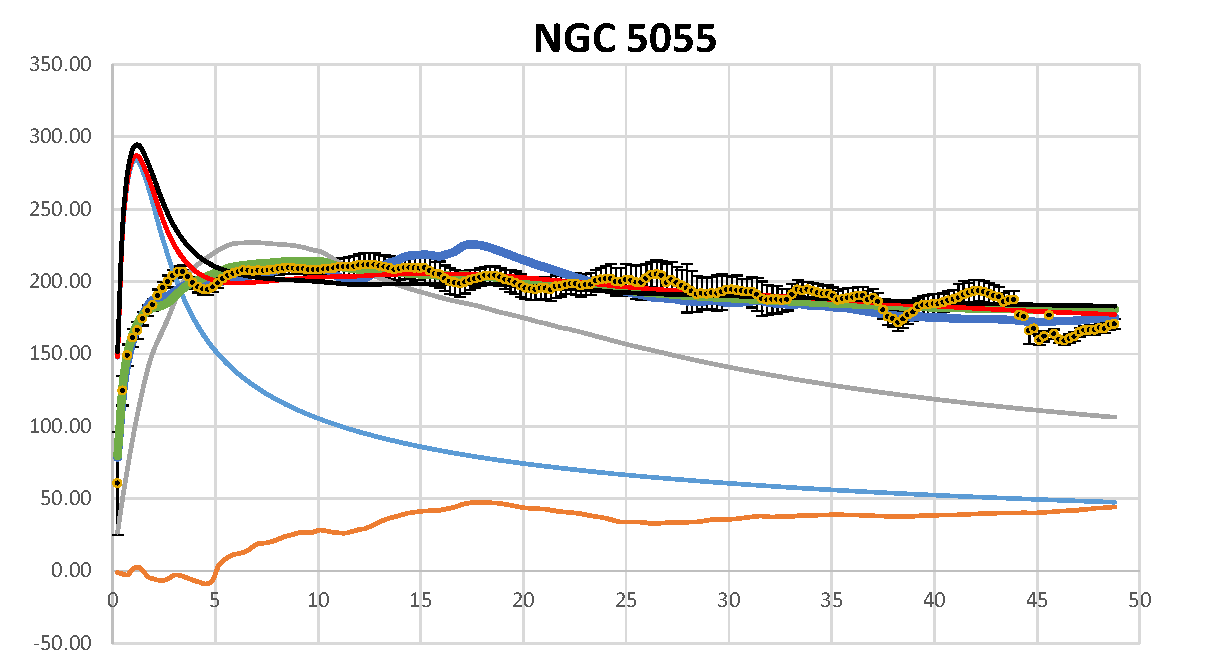}
\end{tabular}
\caption{Graphs of our GR ansatz fits (thick blue), MOND fits (thick green), Burkett DM halo fits (red), and NFW DM halo fits (black) of THINGS galactic RC's (black dots with error bars). Disk contribution is grey, gas contribution is burnt orange, and bulge contribution (when available) is light blue. Vertical axis is rotation velocity in km/s and horizontal axis is orbital radius in kpc. Where the fits are crowded they all conform nicely to the data, so aberrant fitting regions are visible. Note: The MOND distance fitting factor $\gamma_{dist}$ would alter the horizontal scale proportionally. Corresponding numerical results are in Tables \ref{GRRC}, \ref{mondRC}, \ref{BurkettRC}, \& \ref{NFWRC}.} \label{galaxy1}
\end{figure}

\begin{figure}
\begin{tabular}{ll}
\includegraphics[width=3.25in,height=2.5in]{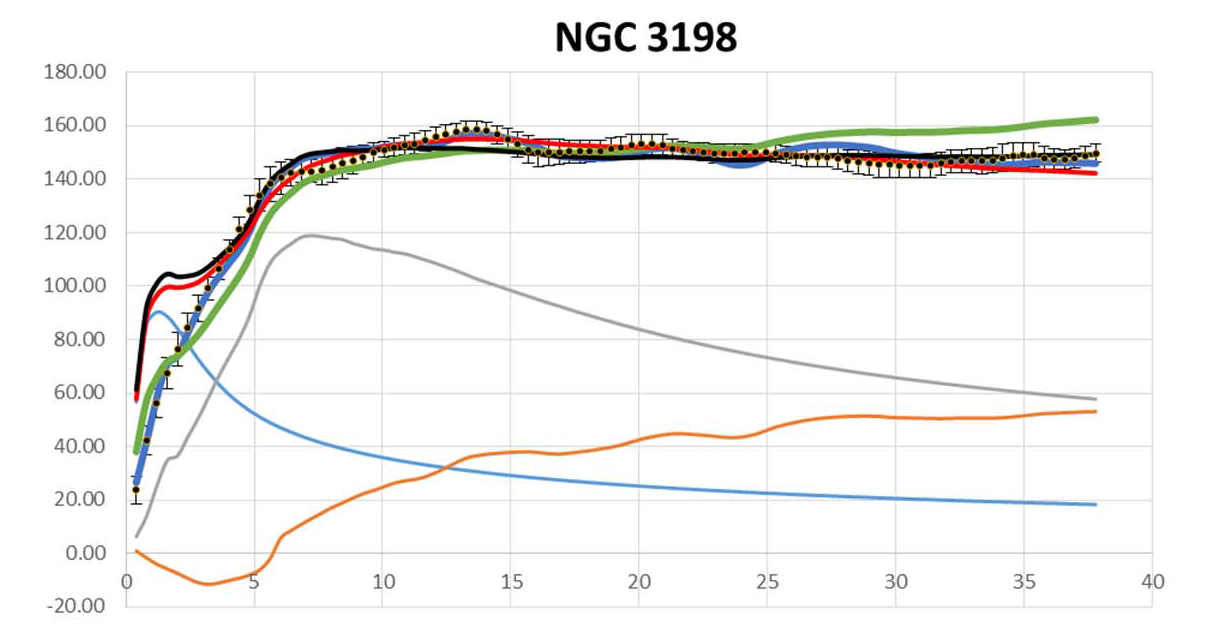} &
\includegraphics[width=3.25in,height=2.5in]{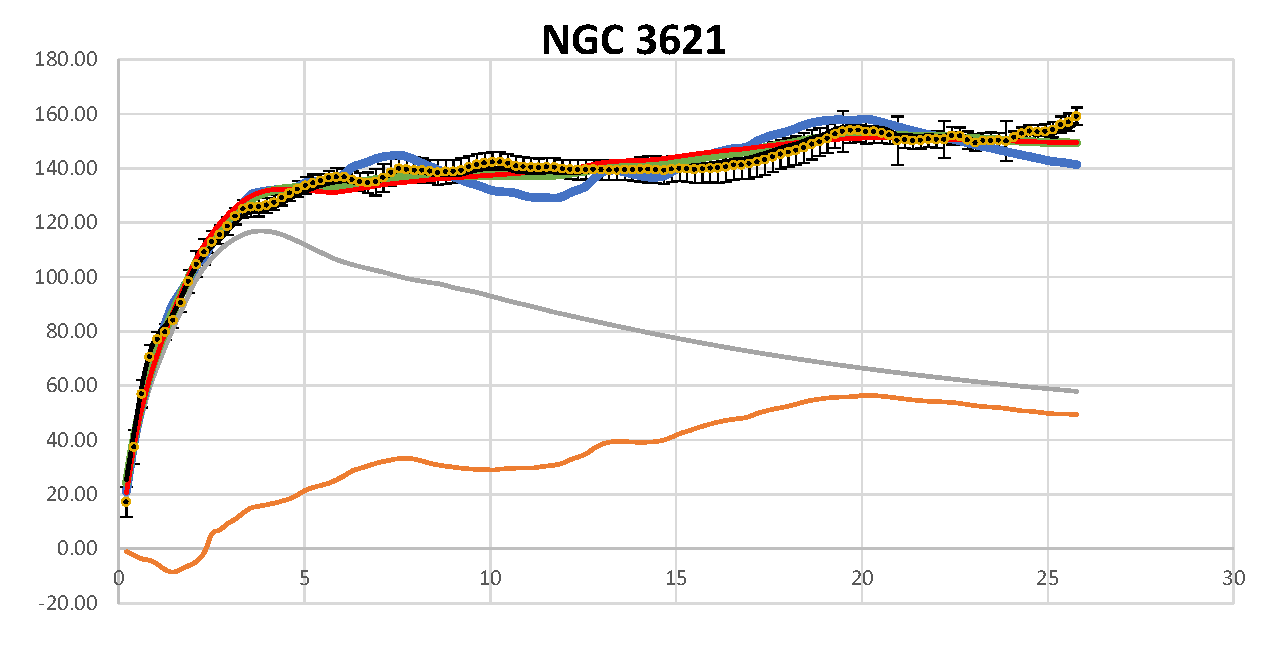} \\
\includegraphics[width=3.25in,height=2.5in]{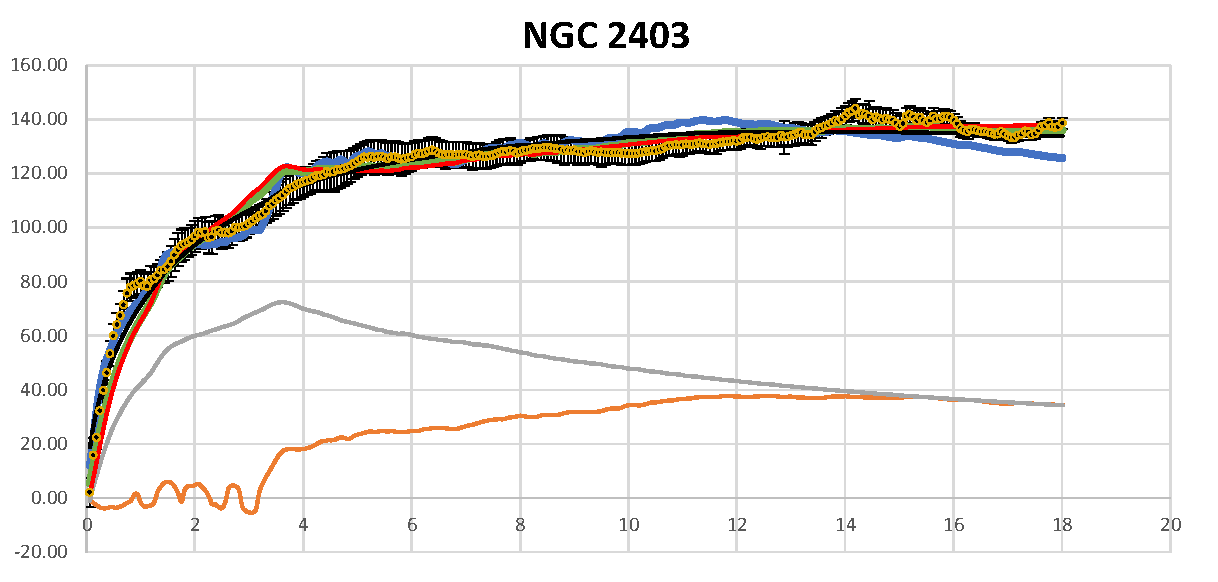} &
\includegraphics[width=3.25in,height=2.5in]{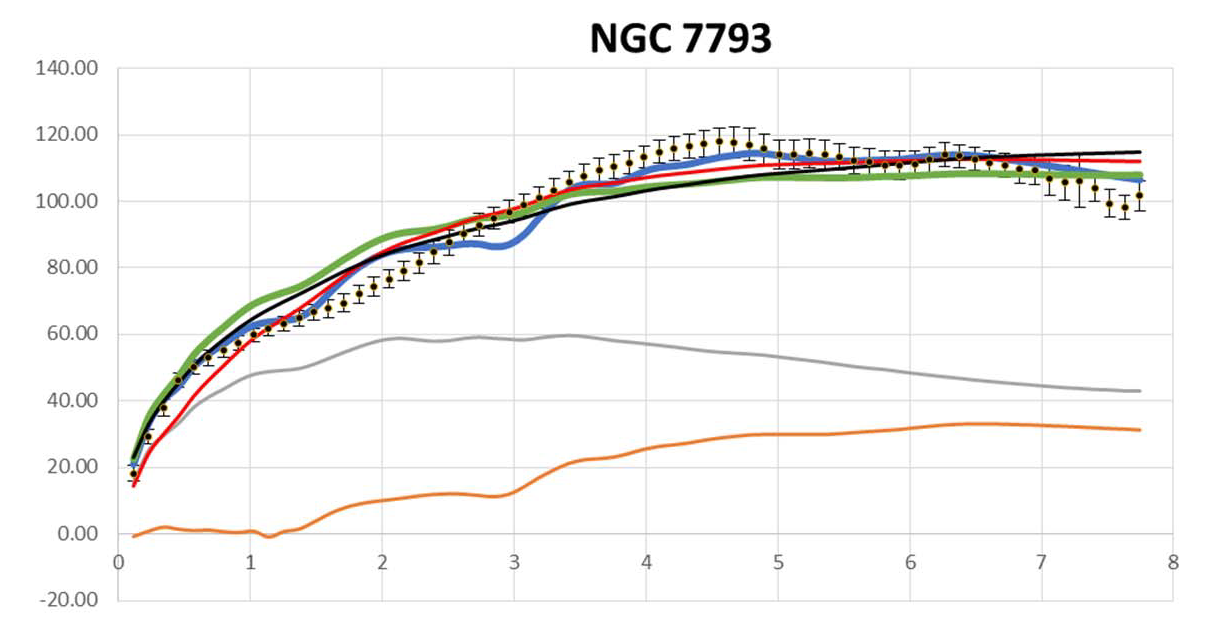} \\
\includegraphics[width=3.25in,height=2.5in]{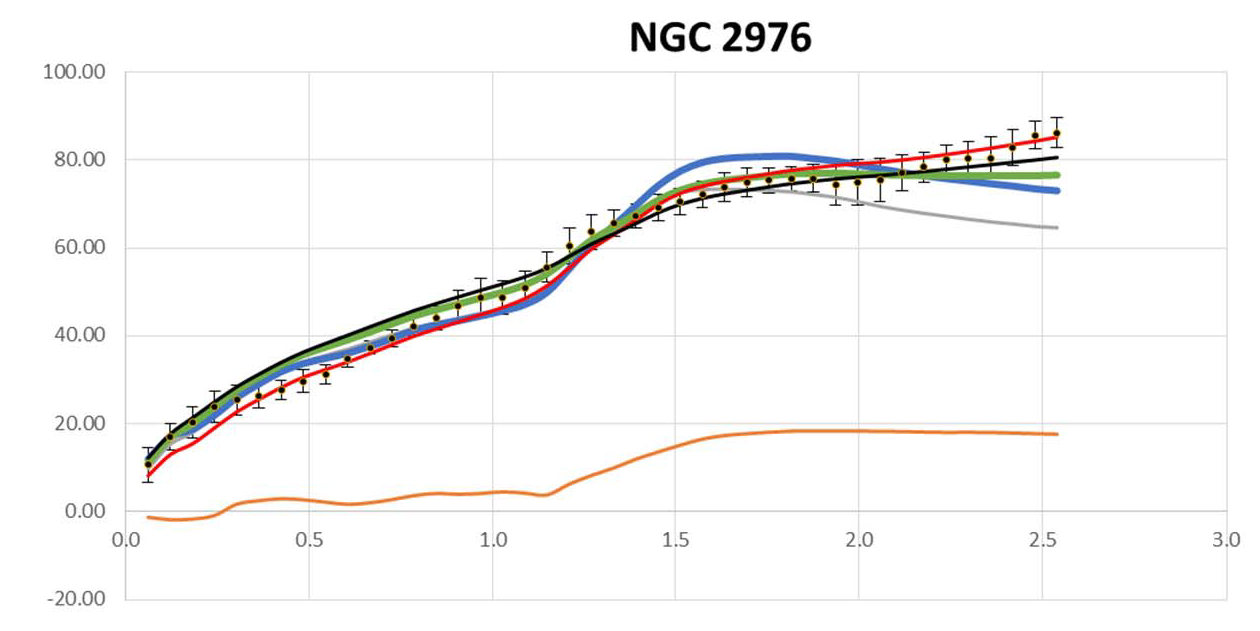} &
\includegraphics[width=3.25in,height=2.5in]{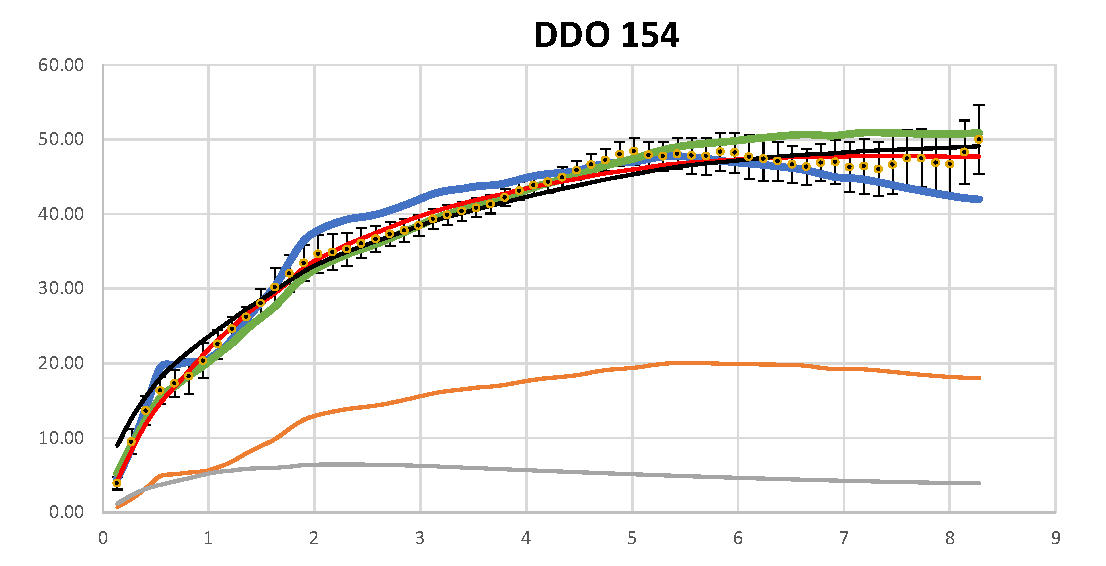}
\end{tabular}
\caption{More graphs of our GR ansatz fits (thick blue), MOND fits (thick green), Burkett DM halo fits (red), and NFW DM halo fits (black) of THINGS galactic RC's (black dots with error bars). Disk contribution is grey, gas contribution is burnt orange, and bulge contribution (when available) is light blue. Vertical axis is rotation velocity in km/s and horizontal axis is orbital radius in kpc. Where the fits are crowded they all conform nicely to the data, so aberrant fitting regions are visible. Note: The MOND distance fitting factor $\gamma_{dist}$ would alter the horizontal scale proportionally. Corresponding numerical results are in Tables \ref{GRRC}, \ref{mondRC}, \ref{BurkettRC}, \& \ref{NFWRC}.}  \label{galaxy2}
\end{figure}

Continuing with an analysis of THINGS fits, $A_{disk}$ for Eq(\ref{GRansatz}) ($\overline{A}_{disk} = (4.4 \pm 7.4$) kpc) makes no contribution in seven of twelve galaxies and is only a small fraction of $A_{gas}$ ($\overline{A}_{gas} = (2370 \pm 4910)$ kpc) when it's not zero. By comparison, Eq(\ref{GRansatz1}) fits have some non-negligible values for $A_{disk}$, e.g., for DDO 154 $A_{disk}$ = 9.18 kpc while $A_{gas}$ = 5.01 kpc. This is consistent with the Burkett DM halo fit for DDO 154 in Table \ref{BurkettRC} where we found $\Upsilon_{\ast}^{3.6}$ = 5.25 and Siutsou et al. found $\Upsilon_{\ast}^{3.6}$ = 1.06 \cite{siutsou}, which increases the disk contribution by a factor of two over its THINGS value. As with bulge data, disk data are non-negligible for small $r$ where Eq(\ref{GRansatz}) is large yet missing mass is small. Thus, if Eq(\ref{GRansatz}) is accurate, then it must be the case that the gas matter is largely responsible for the missing mass while the bulge and disk matter do not contribute appreciably. The trace of our GR ansatz fit with the gas distribution is particularly pronounced in NGC 2841, NGC 2903, and NGC 5055 in Figure \ref{galaxy1}, and NGC 3621, NGC 7793, and NGC 2976 in Figure \ref{galaxy2}. Accentuating the gas contribution while ignoring the bulge and downplaying the disk contributions is akin to a cored DM halo contribution. We found $B$ for Eq(\ref{GRansatz}) fits is small to nonexistent in all galaxies ($\overline{B} = 0.54 \pm 0.81$). In fact, $B$ is zero in fits of NGC 2841 and NGC 2903 where $A_{bulge}$ is also zero, and $B$ and $A_{disk}$ are both zero in DDO 154, so Eq(\ref{GRansatz}) says the bulge and disk, respectively, contribute \textit{absolutely nothing} to the rotational dynamics in those galaxies. This only happens for NGC 2903 bulge data in Eq(\ref{GRansatz1}) fits, but it does happen. Clearly this is an artifact of the simplistic functional forms of these GR ansatzes and should not be taken literally.

\begin{figure}
\begin{center}
\includegraphics [height = 70mm]{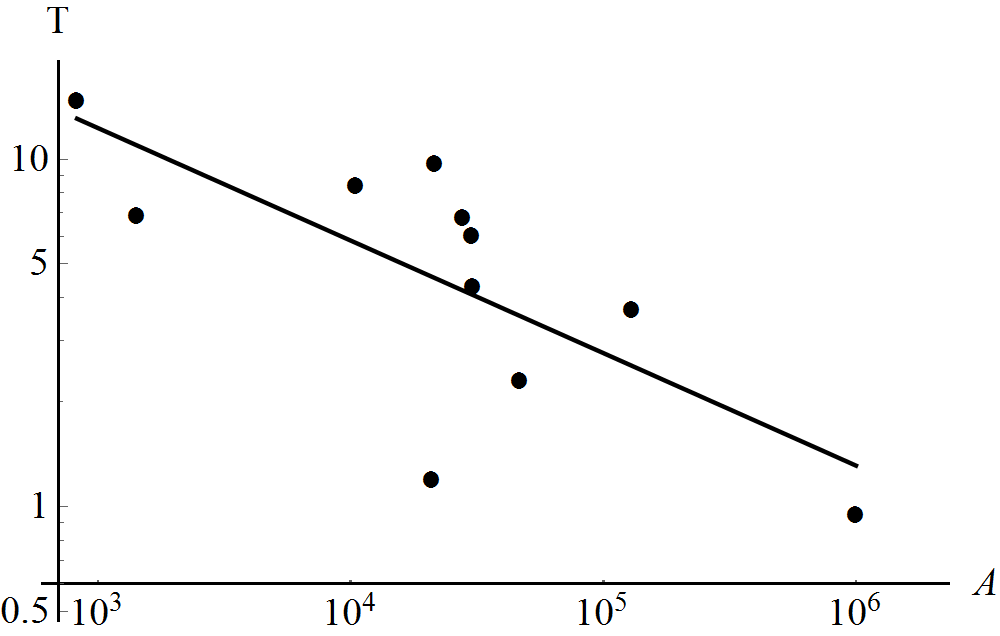}  \caption{This is a log-log plot of X-ray cluster temperature in keV versus $A$ in kpc. There is a trend downward with a p-value $= 0.0118$ and $R = -0.724$.} \label{Atemp}
\end{center}
\end{figure}

\begin{figure}
\begin{center}
\includegraphics [height = 70mm]{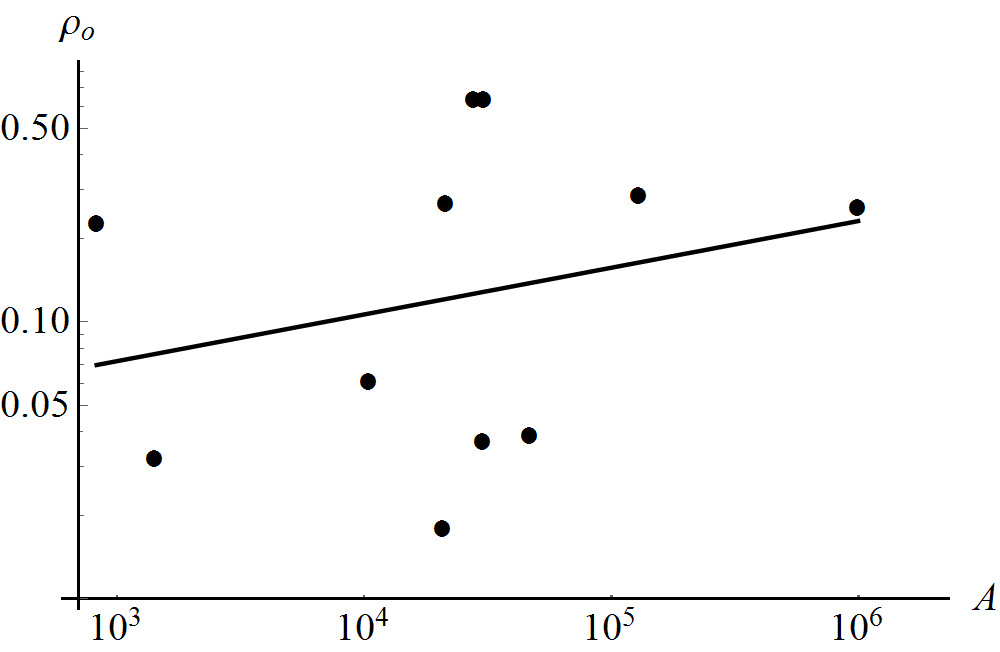}  \caption{This is a log-log plot of X-ray cluster best-fitting King $\beta$-model central density $\rho_o$ in $10^{-25}$ g cm$^{-3}$ versus $A$ in kpc. There is no correlation evident with a p-value of 0.450 and $R = 0.255$.} \label{Adensity}
\end{center}
\end{figure}

\begin{figure}
\begin{center}
\includegraphics [height = 70mm]{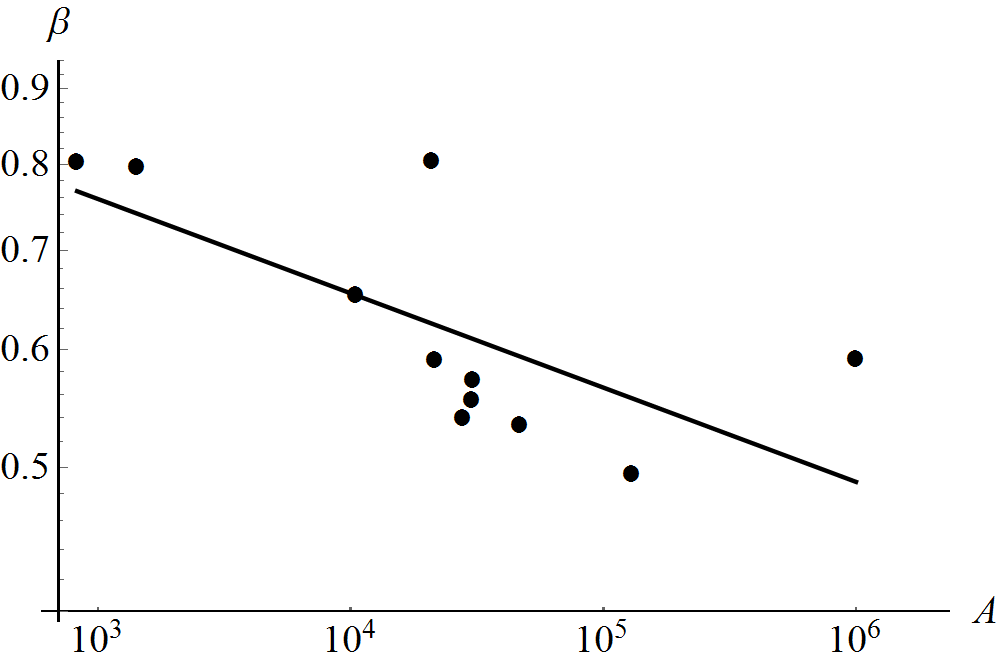}  \caption{This is a log-log plot of X-ray cluster best-fitting King $\beta$-model $\beta$ (dimensionless) versus $A$ in kpc. There is a trend downward with a p-value of 0.0185 and $R = -0.691$.} \label{Abeta}
\end{center}
\end{figure}

\begin{figure}
\begin{center}
\includegraphics [height = 70mm]{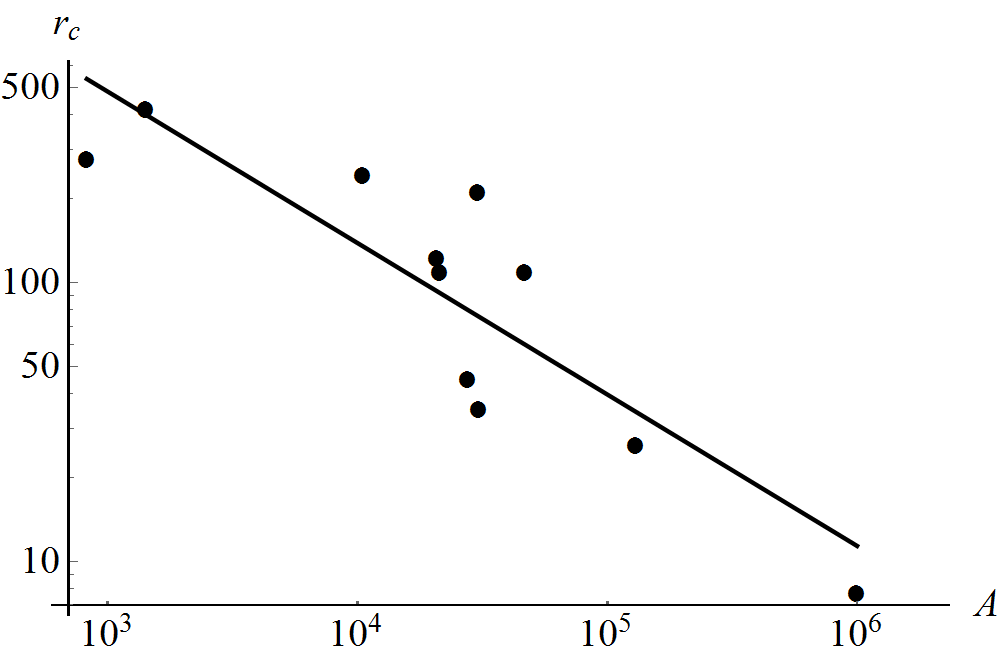}  \caption{This is a log-log plot of X-ray cluster best-fitting King $\beta$-model core radius $r_c$ in kpc versus $A$ in kpc. There is a strong trend downward with a p-value of 0.000438 and $R =-0.874$.} \label{Acore}
\end{center}
\end{figure}

\begin{figure}
\begin{center}
\includegraphics [height = 70mm]{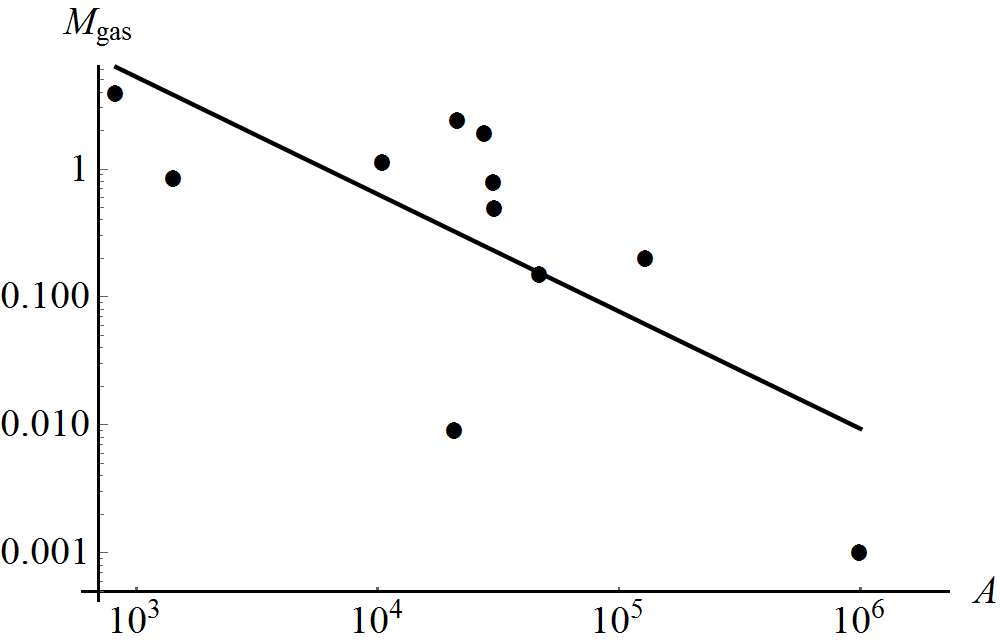}  \caption{This is a log-log plot of X-ray cluster total ICM gas mass $M_{gas}$ in $10^{14}$ $\mbox{M}_{\odot}$ according to Eq(\ref{Eq18}) versus $A$ in kpc. There is a trend downward with a p-value of 0.0161 and $R = -0.702$.} \label{Agas}
\end{center}
\end{figure}

\begin{figure}
\begin{center}
\includegraphics [height = 70mm]{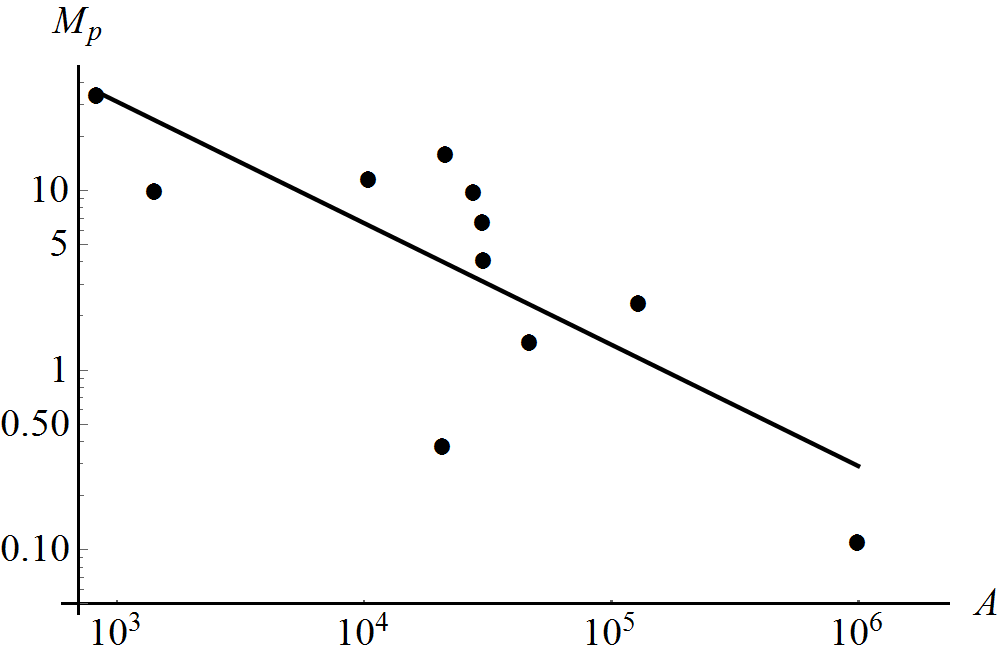}  \caption{This is a log-log plot of X-ray cluster total proper mass $M_p$ according to Eq(\ref{Eq20}) in $10^{14}$ $\mbox{M}_{\odot}$ versus $A$ in kpc. There is a strong trend downward with a p-value of 0.00639 and $R = -0.762$.} \label{Aproper}
\end{center}
\end{figure}

As with galactic RC's, the GR ansatz Eq(\ref{GRansatz}) fits of X-ray cluster mass profiles (ROSAT/ASCA data, average MSE = 0.00535, Table \ref{GRclusters}) compare well with MSTG (average MSE = 0.0236, Table \ref{mstgclusters}) and core-modified NFW DM (average MSE = 0.00975, Table \ref{DMclusters}). It works well for these data because it makes larger corrections for smaller $r$, which is precisely why the GR ansatz Eq(\ref{GRansatz1}) fails for X-ray clusters, i.e., it makes smaller corrections for smaller $r$. The best fit values for $A$ in our GR ansatz Eq(\ref{GRansatz}) fits of X-ray cluster mass profiles (Table \ref{GRclusters}) are not totally surprising, given the exact correction function Eq(\ref{exactclustercorrection}) is a function of temperature, core radius, and $\beta$ and we find our best fit values of $A$ are correlated with all these variables. Specifically, the p-value for a log-log plot of temperature versus $A$ is 0.0118 (Figure \ref{Atemp}), for the best-fitting King $\beta$-model $\beta$ versus $A$ it's 0.0.0185 (Figure \ref{Abeta}), and for the best-fitting King $\beta$-model core radius versus $A$ it's 0.000438 (Figure \ref{Acore}). However, the exact correction function Eq(\ref{exactclustercorrection}) is also a function of the best-fitting King $\beta$-model central density $\rho_o$ which did not correlate with $A$ (p-value of 0.450, Figure \ref{Adensity}). In addition, we found two other correlations in log-log plots of X-ray cluster properties from Table 5.1 of Brownstein \cite{brownstein} versus the best fit values of $A$, i.e., for $A$ and the total mass of ICM gas according to Eq(\ref{Eq18}) the p-value is 0.0161 (Figure \ref{Agas}), and for $A$ and the total proper mass according to Eq(\ref{Eq20}) the p-value is 0.00639 (Figure \ref{Aproper}). These many correlations likely go hand in hand with the overall efficacy of Eq(\ref{GRansatz}) in fitting the X-ray cluster data, an efficacy that is somewhat surprising given the far more complex Eq(\ref{exactclustercorrection}) required to make the fits exact.

\begin{figure}
\begin{center}
\includegraphics [height = 70mm]{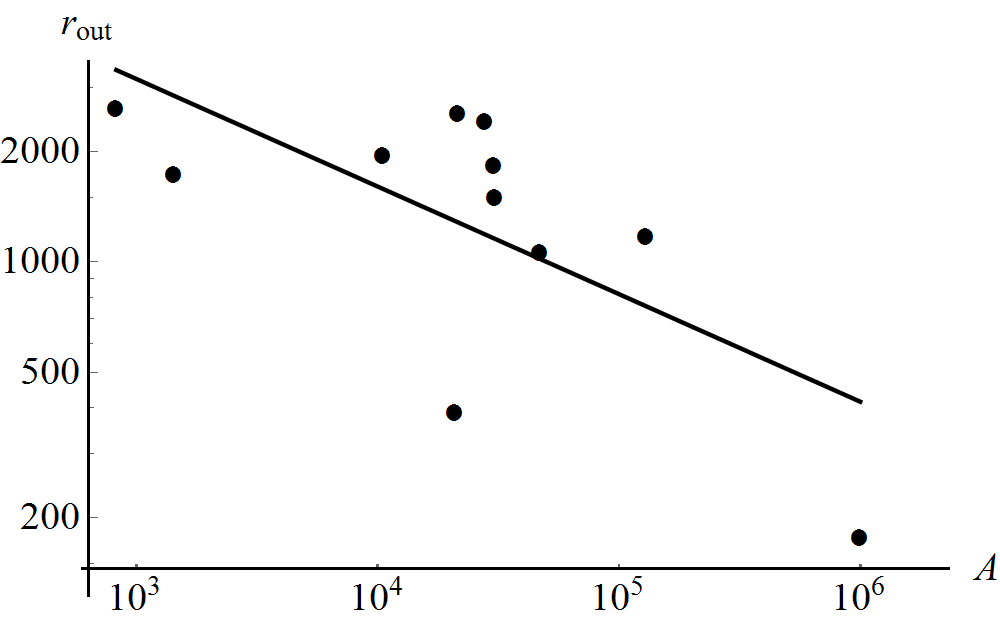}  \caption{This is a log-log plot of $r_{out}$ in kpc, the outer most radial value used for X-ray cluster mass profiles, versus $A$ in kpc. There is a slight trend downward with a p-value of 0.0243 and $R = -0.669$.} \label{Aclusters}
\end{center}
\end{figure}

\begin{figure}
\begin{center}
\includegraphics [height = 70mm]{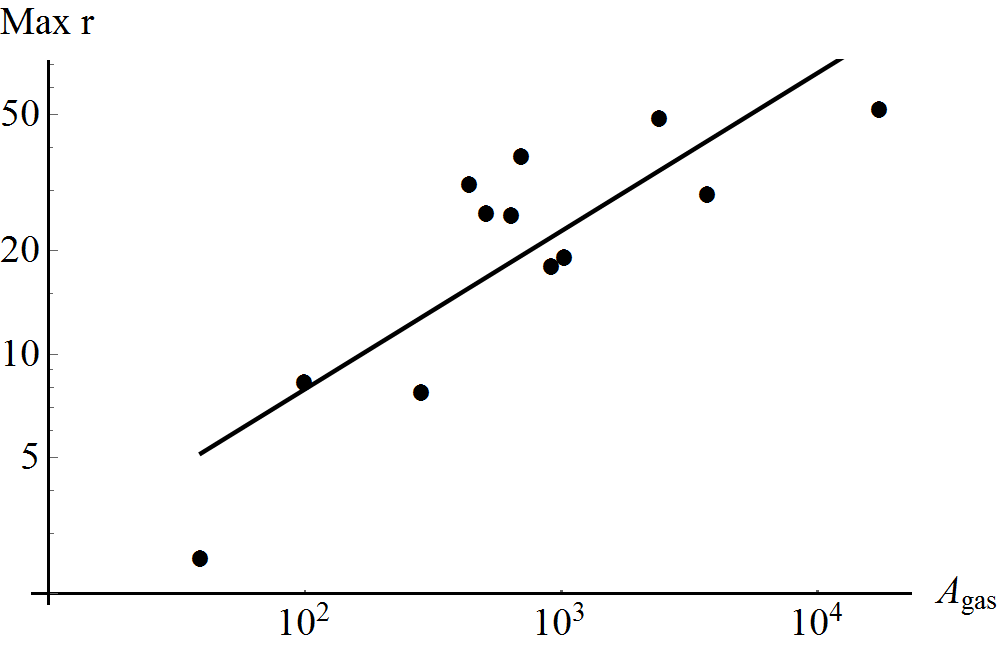}  \caption{This is a log-log plot of maximum radial value of THINGS data in kpc versus $A_{gas}$ in kpc. Contrary to $r_{out}$ versus $A$ for X-ray clusters, there is a strong trend upward with a p-value of 0.000653 and $R = 0.839$.} \label{Agalaxies}
\end{center}
\end{figure}

\begin{figure}
\begin{center}
\includegraphics [height = 70mm]{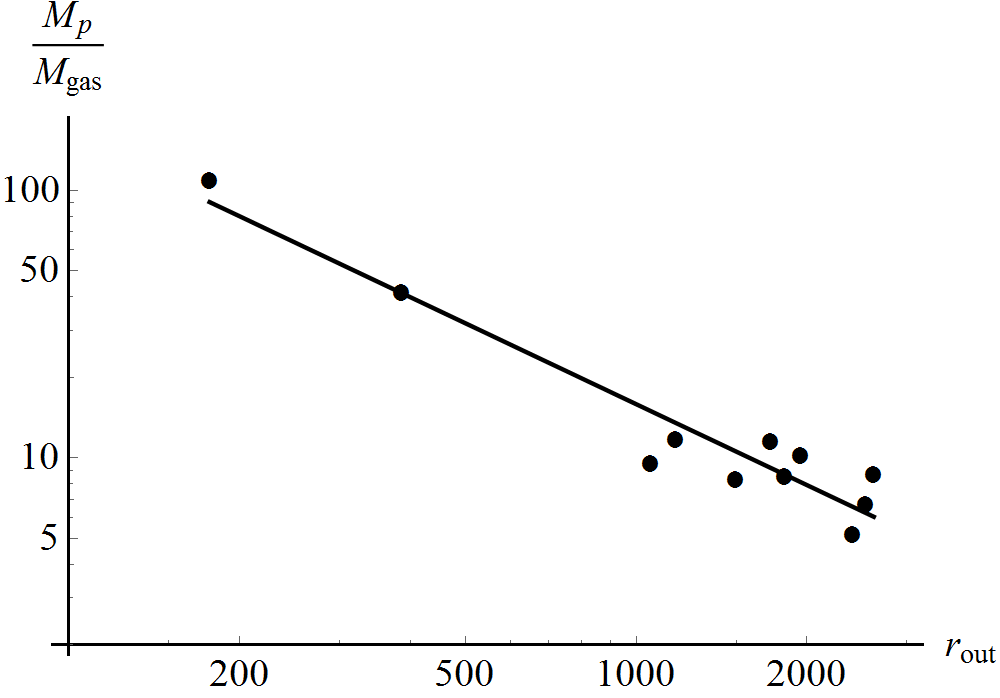}  \caption{This is a log-log plot of the ratio of X-ray cluster total proper mass to total gas mass $\frac{M_p}{M_{gas}}$ (see Figures \ref{Agas} \& \ref{Aproper}) versus $r_{out}$ in kpc. There is a strong trend downward with a p-value of 0.00000289 and $R = -0.960$.} \label{RoutMass}
\end{center}
\end{figure}

\begin{figure}
\begin{center}
\includegraphics [height = 70mm]{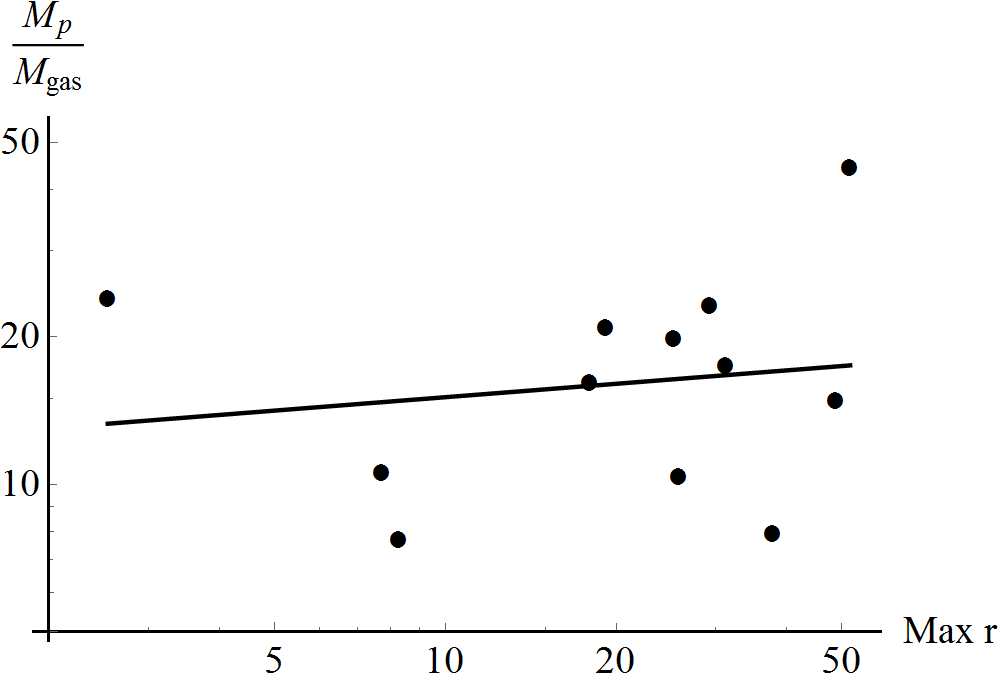}  \caption{This is a log-log plot of the ratio of galaxy total proper mass to total gas mass $\frac{M_p}{M_{gas}}$ versus maximum radial value of THINGS data in kpc. There is no correlation evident with a p-value of 0.626 and $R = 0.157$.} \label{MaxRmass}
\end{center}
\end{figure}

There is an interesting relationship between average $A_{gas}$ in galaxies and average $A$ in X-ray clusters. Again, the X-ray cluster mass profile fits ignored the mass of the galaxies and focused on the ICM gas which is believed to constitute 90\% of the total baryonic cluster mass. For X-ray clusters $\overline{A} = (1.20 \pm 2.94) \times 10^5$ kpc (Table \ref{GRclusters}). Thus, $\overline{A}$ is $\sim 100$ times greater than $\overline{A}_{gas}$ for the galactic RC's (see above). This aligns roughly with the fact that the smallest outer radial data point for clusters (Messier 49) is 175 kpc while the smallest outer radial data point for galaxies (NGC 2976) was 2.54 kpc. And, the largest outer radial data point for clusters (Bullet) is 2620 kpc while the largest outer radial data point for galaxies (NGC 2841) was 51.7 kpc. This is consistent with the log-log correlation between $A_{gas}$ in galaxies and max radial data point which is strongly upward with a p-value of 0.000653 and R = 0.839 (Figure \ref{Agalaxies}). However, within X-ray clusters, the log-log correlation between $A$ and size ($r_{out}$) actually shows a slight \textit{downward} trend with a p-value of 0.0243 and R = -0.669 (Figure \ref{Aclusters}). This corresponds to a log-log plot of total proper-mass-to-gas-mass ratio versus $r_{out}$ for X-ray clusters where we find a strong downward trend with a p-value of $2.89 \times 10^{-6}$ and R = -0.960 (Figure \ref{RoutMass}). But, when we look at a log-log plot of total proper-mass-to-gas-mass ratio versus max radial data point for galaxies there is no correlation evident, p-value = 0.626 and R = 0.157 (Figure \ref{MaxRmass}). Adding total bulge and disk mass to total gas mass for this plot is even less correlated with a p-value of 0.750. 

Finally, we do see an increased contribution from the constant $B$ ($\overline{B} = 600 \pm 1100$) in X-ray cluster fits over the value from Eq(\ref{GRansatz}) obtained for galactic RC's, i.e., $\displaystyle \frac{\overline{B}}{\overline{A}_{gas}} = 0.00023 \: \mbox{kpc}^{-1}$ for galactic RC's while $\displaystyle \frac{\overline{B}}{\overline{A}} = 0.0050 \: \mbox{kpc}^{-1}$ for X-ray clusters. This trend continues to the angular power spectrum of the CMB where $A = 0$ and the correction is entirely in $B$. As with galactic RC's and X-ray cluster mass profiles, the GR ansatz Eq(\ref{GRansatz}) worked well for fitting the first three peaks of the CMB angular power spectrum without DM. The root-mean-square error (RMSE) for the HuS sCDM fit points used is 225 $\left(\mu K \right)^2$. The STVG best fit to these data would equal the $\Lambda$CDM best fit and the RMSE for the $\Lambda$CDM fit corresponding to the HuS fit points used is 240 $\left(\mu K \right)^2$, although that fit was made for all $\ell$ in the range $30 \leq \ell \leq 2508$ (Figure \ref{PlanckDataFit}).

\section{Conclusions}\label{sec7}
Twelve galactic RC's, eleven X-ray cluster mass profiles and the first three peaks of the CMB angular power spectrum do not provide enough data fits to draw any strong conclusions per se, but the general trends and results noted above are consistent with well-established research. Therefore, we believe it's not unreasonable to draw some speculative conclusions. 

First, since ``missing mass'' doesn't necessarily entail ``missing matter'' in GR, it is reasonable to ask whether or not this GR contextuality can resolve the missing mass problem without invoking DM. Therefore, we explored various GR ansatzes for fitting galactic RC's, X-ray cluster mass profiles, and the angular power spectrum of the CMB without DM. Eqs(\ref{GRacceleration} \& \ref{exactclustercorrection}) do not bode well for the prospect of finding an exact GR solution that will produce a simple functional form for $f(r)$ in obtaining proper mass from dynamic mass $dM_p = f(r)dM$ for realistic astrophysical situations. Indeed, it seems highly unlikely an exact GR solution that in anyway corresponds to the astrophysical matter distributions of galaxies, clusters, voids, filaments, mergers, etc. will ever be produced. The strength of GR lies in its versatility, i.e., any self-consistent metric and stress-energy tensor on the spacetime manifold per Einstein's equations is a valid solution. But, this is also its weakness, i.e., Einstein's equations are notoriously complex. The success of the simple FLRW cosmology models on large scales should not be expected for galactic and cluster scales. A GR ansatz based on the integrated difference in spatial geometry between two spacetime contexts for one and the same matter Eq(\ref{integratedmassratio}) was found to provide fits for galactic RC's comparable to MOND, Burkett DM halo, and NFW DM halo fits, but failed to provide a reasonable fit for the mass profiles of X-ray clusters. However, since these scales are dominated by weak gravitational fields, it is perhaps not surprising that simple GR perturbation theory exploiting spatial symmetry has produced a simple GR ansatz Eq(\ref{GRansatz}) that compared favorably with MOND, MSTG, STVG, Burkett halo DM, NFW halo and core-modified DM, and $\Lambda$CDM fits of galactic RC's, X-ray cluster mass profiles and the CMB angular power spectrum. The same idea was used in the case of dark energy where perturbative metric corrections to proper distance in the flat space of Einstein-de Sitter (EdS) cosmology produced a fit of the SCP Union2.1 SN Ia data without dark energy (no $\Lambda$) equal to that of $\Lambda$CDM. It may well be that we will have to settle for an empirical \cite{mcgaughDM} fit of data on these scales as motivated by GR-allowable concepts, rather than exact GR solutions. Certainly this approach is no less motivated than DM fits of the data, since the fitting parameters there are not specified by particle physics. 

Second, the best fit values of the fitting parameters for Eq(\ref{GRansatz}) showed interesting trends across and within the three data sets lending to the viability of the approach. The average ratio of total proper mass to total dynamic mass in the THINGS data was $4.19 \pm 0.81$. This is consistent with a dark matter fraction of 79\% in galaxies found using microlensing (\cite{jimenez} and references therein). As we showed in Figure \ref{MaxRmass}, this ratio is uncorrelated with maximum radial value of THINGS data (p-value of 0.626 and $R = 0.157$). This is quite different than the strong trend downward found in a log-log plot of the ratio of X-ray cluster total proper mass to total gas mass versus $r_{out}$ (p-value of 0.00000289 and $R = -0.960$, Figure \ref{RoutMass}). Therefore, we might expect a different GR ansatz will work better at galactic scales than at cluster scales, as we found. 

Clearly, more data fits on these scales and others are needed to establish the viability of this approach. Obviously, we would prefer a more compelling motivation for the GR ansatz and a correspondingly more robust heuristic in the absence of an exact solution. Perhaps a numerical approach with an algorithm that mirrors connected exact solutions could produce a more justifiable GR ansatz. Regge calculus \cite{wong,regge,reggeMTW,barrett,williams} might be our best numerical bet for dealing with the compatibility of the extrinsic curvature and 3D metric at the interface between disparate spacetimes, given its inherent (3 + 1) formulation. In conclusion, while this approach is nascent, we believe it is worth exploring.

\section*{Acknowledgements}
We thank Erwin de Blok for providing the THINGS data, Gianfranco Gentile for providing information about his MOND fits, Joel Brownstein for providing information about MSTG, Thomas Reiprich for providing information about X-ray cluster mass data, Stacy McGaugh for providing input about MOND as relates to cosmology, Wayne Hu for help with HuS, and Viktor Toth for help with the STVG fit of Planck 2015.

\vspace{6pt} 




\renewcommand\bibname{References}


\end{document}